%% file: Main.tex
\documentclass[journal,12pt,onecolumn]{IEEEtran}

\usepackage{setspace}
\ifCLASSOPTIONonecolumn \doublespace \fi
\usepackage{graphics}
\usepackage{rotating}
\usepackage{amsfonts}
\usepackage{amsmath}
\usepackage{amsthm}
\usepackage{subfigure}
\usepackage{multirow}
\allowdisplaybreaks[1]
\newtheorem*{theorem}{Theorem}
\renewcommand{\IEEEQED}{\IEEEQEDopen}

\graphicspath{{Figures/}} \DeclareGraphicsExtensions{.eps,.ps}

\begin{document}

\title{Multiple Beamforming with Perfect Coding}

\ifCLASSOPTIONconference
\author{\IEEEauthorblockN{Boyu Li and Ender Ayanoglu}\\
\IEEEauthorblockA{Center for Pervasive Communications and Computing\\
Department of Electrical Engineering and Computer Science\\
University of California, Irvine\\
Email: boyul@uci.edu, ayanoglu@uci.edu}} \else
\author{Boyu~Li,~\IEEEmembership{Student~Member,~IEEE,}~and~Ender~Ayanoglu,~\IEEEmembership{Fellow,~IEEE}
\thanks{B. Li and E. Ayanoglu are with the Center for Pervasive Communications and Computing, Department of Electrical Engineering and Computer Science, Henry Samueli School of Engineering, University of California, Irvine, CA 92697-3975 USA (e-mail: boyul@uci.edu; ayanoglu@uci.edu).}} \fi

\maketitle

\ifCLASSOPTIONonecolumn
 \setlength\arraycolsep{4pt}
\else
 \setlength\arraycolsep{2pt}
 \def\sizefig{0.95}
\fi

\input{Abstract}


\input{Introduction}

\input{System_model}
\input{PCMB}
\input{BICMB-PC}
\input{Results}
\input{Conclusion}
\input{Acknowledgement}
\input{Appendix}

\bibliographystyle{IEEEtran}
\bibliography{IEEEabrv,Mybib}

\end{document}

%% file: Abstract.tex
\begin{abstract}

Perfect Space-Time Block Codes (PSTBCs) achieve full diversity, full rate, nonvanishing constant minimum determinant, uniform average transmitted energy per antenna, and good shaping. However, the high decoding complexity is a critical issue for practice. When the Channel State Information (CSI) is available at both the transmitter and the receiver, Singular Value Decomposition (SVD) is commonly applied for a Multiple-Input Multiple-Output (MIMO) system to enhance the throughput or the performance. In this paper, two novel techniques, Perfect Coded Multiple Beamforming (PCMB) and Bit-Interleaved Coded Multiple Beamforming with Perfect Coding (BICMB-PC), are proposed, employing both PSTBCs and SVD with and without channel coding, respectively. With CSI at the transmitter (CSIT), the decoding complexity of PCMB is substantially reduced compared to a MIMO system employing PSTBC, providing a new prospect of CSIT. Especially, because of the special property of the generation matrices, PCMB provides much lower decoding complexity than the state-of-the-art SVD-based uncoded technique in dimensions $2$ and $4$. Similarly, the decoding complexity of BICMB-PC is much lower than the state-of-the-art SVD-based coded technique in these two dimensions, and the complexity gain is greater than the uncoded case. Moreover, these aforementioned complexity reductions are achieved with only negligible or modest loss in performance.  


\end{abstract}

%% file: Introduction.tex
\section{Introduction} \label{sec:Introduction}

In a Multiple-Input Multiple-Output (MIMO) system, when the Channel State Information (CSI) is available at the transmitter as well as the receiver, beamforming techniques, which exploit Singular Value Decomposition (SVD), are applied to achieve spatial multiplexing\footnotemark \footnotetext{In this paper, the term ``spatial multiplexing" is used to describe the number of spatial subchannels, as in \cite{Paulraj_ST}. Note that the term is different from ``spatial multiplexing gain" defined in \cite{Zheng_DM}.}  and thereby increase the data rate, or to enhance performance \cite{Jafarkhani_STC}. Nevertheless, spatial multiplexing without channel coding results in the loss of the full diversity order \cite{Sengul_DA_SMB}. To overcome the diversity degradation, Bit-Interleaved Coded Multiple Beamforming (BICMB) interleaving the bit codeword through the multiple subchannels with different diversity orders was proposed \cite{Akay_BICMB}, \cite{Akay_On_BICMB}. BICMB can achieve full diversity as long as the code rate $R_c$ and the number of employed subchannels $S$ satisfy the condition $R_cS \leq 1$ \cite{Park_DA_BICMB}, 
\cite{Park_DA_BICMB_J}. Moreover, by employing the constellation precoding technique, full diversity and full multiplexing can be achieved simultaneously for both uncoded and convolutional coded SVD systems with the trade-off of a higher decoding complexity \cite{Park_CPB}, \cite{Park_BICMB_CP}, \cite{Park_MB_CP}, \cite{Park_CPMB}. Specifically, in the uncoded case, full diversity requires that all streams are precoded, i.e., Fully Precoded Multiple Beamforming (FPMB). On the other hand, for the convolutional coded SVD systems without the condition $R_cS \leq 1$, other than full precoding, i.e., Bit-Interleaved Coded Multiple Beamforming with Full Precoding (BICMB-FP), partial precoding, i.e., Bit-Interleaved Coded Multiple Beamforming with Partial Precoding (BICMB-PP), could also achieve both full diversity and full multiplexing with the properly designed combination of the convolutional code, the bit interleaver, and the constellation precoder.

In MIMO systems, space-time coding can be employed to offer spatial diversity \cite{Jafarkhani_STC}. In \cite{Oggier_PSTBC}, Perfect Space-Time Block Codes (PSTBCs) were introduced for dimensions $2$, $3$, $4$, and $6$. PSTBCs have the properties of full rate, full diversity, uniform average transmitted energy per antenna, good shaping of the constellation, and nonvanishing constant minimum determinant for increasing spectral efficiency which offers high coding gain. In \cite{Elia_PSTBC}, PSTBCs were generalized to any dimension. However, it was proved in \cite{Berhuy_PSTC} that particular PSTBCs, yielding increased coding gain, only exist in dimensions $2$, $3$, $4$, and $6$. Due to the advantages of PSTBCs, the Golden Code (GC), which is the best known PSTBC for MIMO systems with two transmit and two receive antennas \cite{Belfiore_GC}, \cite{Dayal_STC}, has been incorporated into the $802.16$e Worldwide Interoperability for Microwave Access (WiMAX) standard \cite{IEEE_802_16e}.

Despite these advantages, the high decoding complexity of PSTBCs, especially for large dimensions, is a critical issue for practical employments. For the PSTBC of dimension $D \in \{2,3,4,6\}$, since each codeword employs $D^2$ information symbols from an $M$-QAM or $M$-HEX \cite{Forney_EM} constellation, $M^{D^2}$ points are calculated by exhaustive search to achieve the Maximum Likelihood (ML) decoding. Therefore, the decoding complexity is proportional to $M^{D^2}$, denoted by $\mathcal{O}(M^{D^2})$. Sphere Decoding (SD) is an alternative for ML with reduced complexity \cite{Jalden_SD}. While SD reduces the average decoding complexity, the worst-case complexity is still $\mathcal{O}(M^{D^2})$. Several techniques have been proposed to reduce the decoding complexity of PSTBCs. In \cite{Howard_PSTBC}, an approach based on the conditional ML was applied to obtain essentially ML performance with the worst-case complexity of $\mathcal{O}(M^{D(D-1)})$. In \cite{Sinnokrot_FMLD_GC}, \cite{Sinnokrot_GC_FD}, \cite{Sinnokrot_STBC_LMLDC}, the worst-case complexity of PSTBCs was reduced to $\mathcal{O}(M^{(D-0.5)D-0.5})$ without performance degradation. In \cite{Howard_FD_GC}, a decoding technique applying the Diophantine approximation was presented for GC with the complexity of $\mathcal{O}(M^{2})$ and the trade-off of $2$dB performance loss. In \cite{Hu_PSTC}, \cite{Sarkiss_PC_GC}, \cite{Othman_GC}, suboptimal decoders for PSTBCs were discussed.

In this paper, two novel techniques are proposed. The first technique, Perfect Coded Multiple Beamforming (PCMB), combines PSTBCs with multiple beamforming and achieves full diversity, full multiplexing, and full rate simultaneously, in a similar fashion to a MIMO system employing PSTBC and FPMB. With the knowledge of CSI at the transmitter (CSIT), the threaded structure of the PSTBC could be separated at the receiver, and the decoding complexity of PCMB is thereby substantially reduced compared to a MIMO system employing PSTBC and similar to FPMB. This result offers a new prospect of CSIT since it is mostly used to enhance either the performance or the throughput of a MIMO system. Especially, because of the special property of the generation matrices in dimensions $2$ and $4$, the real and the imaginary parts of the received signal can be decoded separately, and therefore PCMB provides much lower decoding complexity than FPMB in these two dimensions. For instance, the worst-case decoding complexity of a MIMO system employing GC, FPMB of dimension $2$, and Golden Coded Multiple Beamforming (GCMB), which is the PCMB of dimension $2$, are $\mathcal{O}(M^{2.5})$, $\mathcal{O}(M)$, and $\mathcal{O}(\sqrt{M})$ respectively. On the other hand, the second technique, Bit-Interleaved Coded Multiple Beamforming with Perfect Coding (BICMB-PC) transmits bit-interleaved codewords of PSTBC through the multiple subchannels. BICMB-PC achieves full diversity and full multiplexing simultaneously, in a similar fashion to BICMB-FP. Because the real and imaginary parts of the received signal can be separated, and only the part corresponding to the coded bit is required to calculate one bit metric for the Viterbi decoder in dimensions $2$ and $4$, which also results from the special property of the generation matrices, BICMB-PC achieves much lower decoding complexity than BICMB-FP, and the complexity reduction from BICMB-FP to BICMB-PC is greater than the reduction from FPMB to PCMB in these two dimensions. For instance, the worst-case complexity for acquiring one bit metric of BICMB-FP of dimension $2$ and Bit-Interleaved Coded Multiple Beamforming with Golden Coding (BICMB-GC), which is the BICMB-PC of dimension $2$, are $\mathcal{O}(M)$ and $\mathcal{O}(\sqrt{M})$ respectively. Since the precoded part of BICMB-PP could be considered as a smaller dimensional BICMB-FP, BICMB-PC of dimensions $2$ and $4$ could be applied to replace the precoded part and reduce the complexity for BICMB-PP. Furthermore, these aforementioned complexity reductions achieved by PCMB and BICMB-PC only cause negligible or modest loss in performance.

The remainder of this paper is organized as follows: In Section \ref{sec:System_model}, the descriptions of PCMB and BICMB-PC are given. In Section \ref{sec:PCMB} and \ref{sec:BICMB-PC}, the diversity analysis and decoding technique of PCMB and BICMB-PC in dimension $2$ are first presented, and then generalized to larger dimensions, respectively. In Section \ref{sec:Results}, performance comparisons of different techniques are carried out. Finally, a conclusion is provided in Section \ref{sec:Conclusion}.

\textbf{Notations:} Bold lower (upper) case letters denote vectors (matrices). The notation $\mathrm{diag}[b_1, \ldots, b_D]$ denotes a diagonal matrix with diagonal entries $b_1, \ldots, b_D$. The notations $\Re(\cdot)$ and $\Im(\cdot)$ denote the real and imaginary parts of a complex number, respectively. The superscripts $(\cdot)^H$, $(\cdot)^T$, $(\cdot)^*$, and $\bar{(\cdot)}$ stand for the conjugate transpose, transpose, complex conjugate, and binary complement, respectively. The notation $\lceil \cdot \rceil$ denotes the ceiling function that maps a real number to the next largest integer. The notations $\mathbb{R}^+$ and $\mathbb{C}$ stand for the set of positive real numbers and the complex numbers, respectively.




%% file: System_model.tex
\section{System Model} \label{sec:System_model}

\ifCLASSOPTIONonecolumn
\begin{figure}[!m]
\centering{ \subfigure[]{\includegraphics[width=1.0\linewidth]{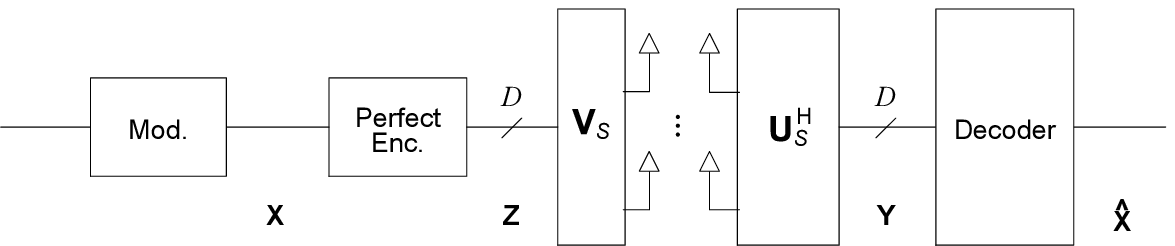} \label{fig:system_model_pcmb}} \hfil
\subfigure[]{\includegraphics[width=1.0\linewidth]{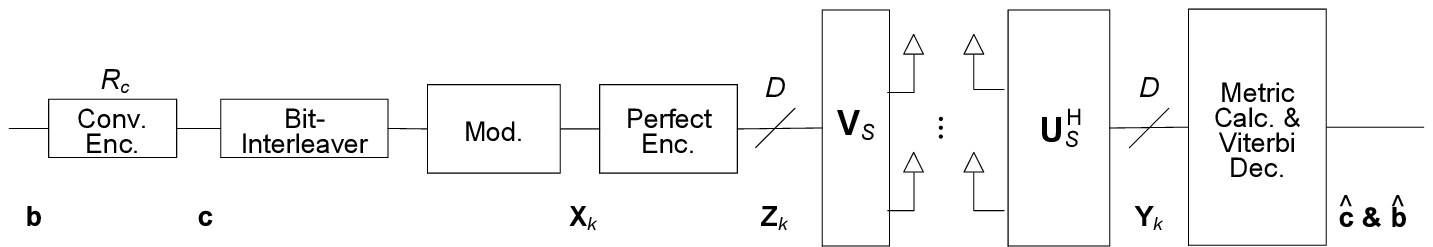} \label{fig:system_model_bicmb-pc}} \hfil
\subfigure[]{\includegraphics[width=1.0\linewidth]{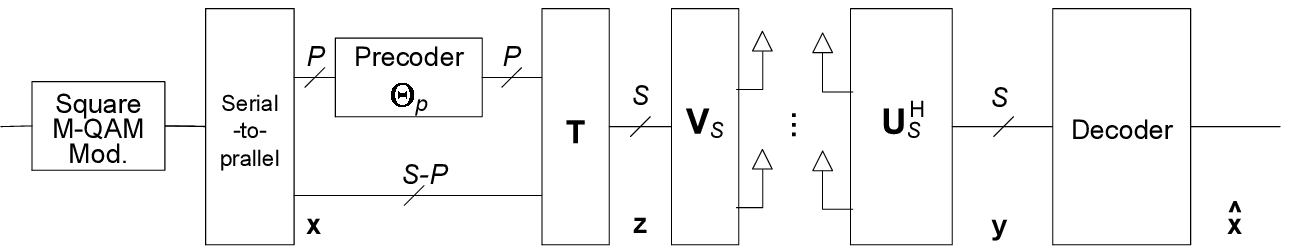} \label{fig:system_model_cpmb}} \hfil
\subfigure[]{\includegraphics[width=1.0\linewidth]{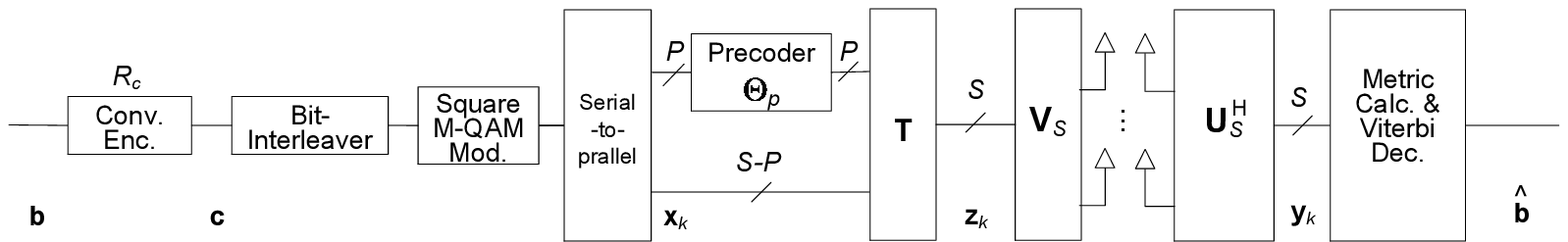} \label{fig:system_model_bicmb-cp}}}
\caption{Structure of (a) PCMB, (b) BICMB-PC, (c) CPMB (FPMB when $P=S$), and (d) BICMB-CP (BICMB-FP when $P=S$).}
\label{fig:system_model}
\end{figure}
\else
\begin{figure}[!t]
\centering{ \subfigure[]{\includegraphics[width=\sizefig\linewidth]{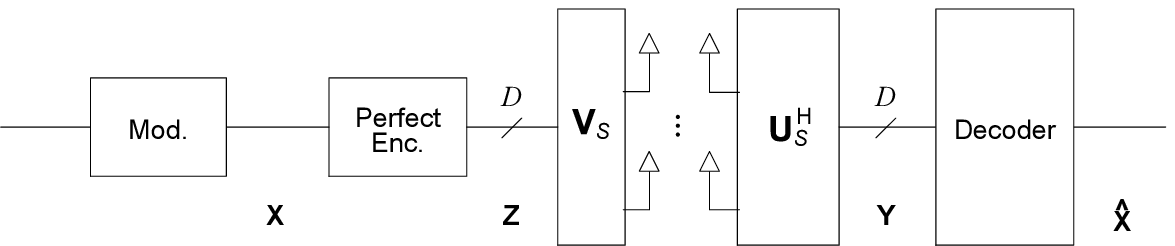} \label{fig:system_model_pcmb}} \hfil
\subfigure[]{\includegraphics[width=\sizefig\linewidth]{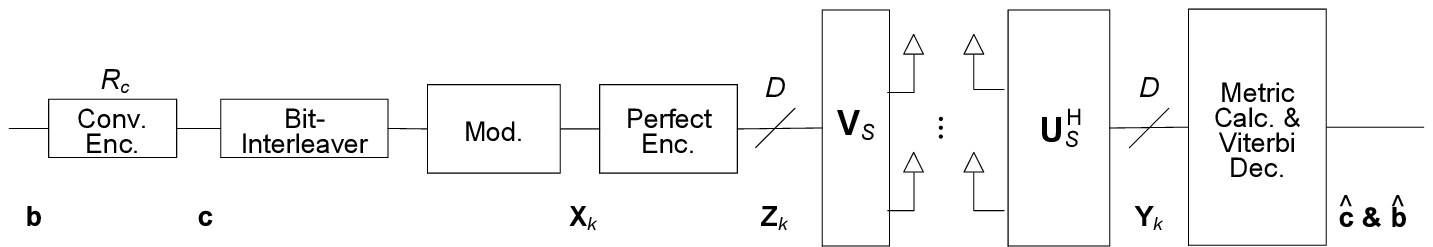} \label{fig:system_model_bicmb-pc}} \hfil
\subfigure[]{\includegraphics[width=\sizefig\linewidth]{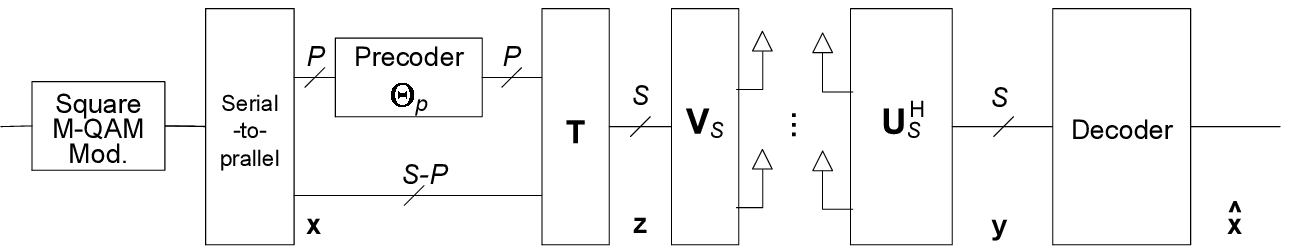} \label{fig:system_model_cpmb}} \hfil
\subfigure[]{\includegraphics[width=\sizefig\linewidth]{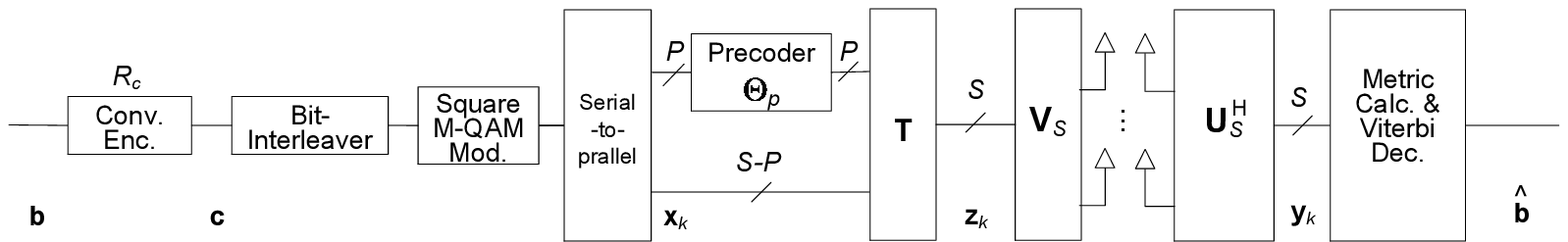} \label{fig:system_model_bicmb-cp}}}
\caption{Structure of (a) PCMB, (b) BICMB-PC, (c) CPMB (FPMB when $P=S$), and (d) BICMB-CP (BICMB-FP when $P=S$).}
\label{fig:system_model}
\end{figure}
\fi

\subsection{PCMB} \label{subsec:PCMB_model}

Fig. \ref{fig:system_model_pcmb} represents the structure of PCMB. The information bit sequence is first mapped by Gray encoding and modulated by $M$-QAM or $M$-HEX
. Then, $D^2$ consecutive complex-valued scalar symbols are encoded into one PSTBC codeword, where $D \in \{2,3,4,6 \}$ is the system dimension. Hence, the PSTBC codeword $\mathbf{Z}$ is constructed as
\begin{equation}
\mathbf{Z}=\sum_{v=1}^D{\mathrm{diag}(\mathbf{G}\mathbf{x}_{v})\mathbf{E}^{v-1}},
\label{eq:PSTBC_pcmb}
\end{equation}
where $\mathbf{G}$ is an $D\times D$ unitary generation matrix, $\mathbf{x}_{v}$ is an $D\times1$ vector whose elements are the $v$th $D$ input scalar symbols, and
\begin{align*}
\mathbf{E} = \left[
\begin{array}{ccccc}
0 & 1 & \cdots & 0 & 0 \\
0 & 0 & 1 & \cdots & \vdots \\
\vdots & \vdots & \ddots & \ddots & \vdots \\
0 & \cdots & \cdots & \cdots & 1 \\
g & 0 & \cdots & 0 & 0
\end{array}
\right],
\end{align*}
with 
\begin{align*}
g = \left\lbrace
\begin{array}{cc}
i, & D=2,4, \\
e^{\frac{2{\pi}i}{3}}, & D=3, \\
-e^{\frac{2{\pi}i}{3}}, & D=6.
\end{array} \right.
\end{align*}
The specific $\mathbf{G}$ matrix for different dimensions can be found in \cite{Oggier_PSTBC}, \cite{Belfiore_GC}.

The MIMO channel $\mathbf{H} \in \mathbb{C}^{N_r \times N_t}$ is assumed to be quasi-static, Rayleigh, and flat fading, and known by both the transmitter and the receiver, where $N_r$ and $N_t$ denote the number of receive and transmit antennas respectively. The beamforming vectors are determined by the SVD of the MIMO channel, i.e., $\mathbf{H} = \mathbf{U \Lambda V}^H$ where $\mathbf{U}$ and $\mathbf{V}$ are unitary matrices, and $\mathbf{\Lambda}$ is a diagonal matrix whose $s$th diagonal element, $\lambda_s \in \mathbb{R}^+$, is a singular value of $\mathbf{H}$ in decreasing order. When $S \leq \min\{N_t,N_r\}$ streams are transmitted at the same time, the first $S$ vectors of $\mathbf{U}$ and $\mathbf{V}$ are chosen to be used as beamforming matrices at the receiver and the transmitter, respectively. For a MIMO system employing PSTBC in dimension $D$, $D^2$ information symbols are transmitted through $D$ time slots. In the case of PCMB, to achieve the same rate as a MIMO system employing PSTBC, the number of streams is $S$ where $N_t=N_r=S=D \in \{2,3,4,6\}$.

The received signal is
\begin{align}
\mathbf{Y} =  \mathbf{U}^{H}\mathbf{HVZ}+\mathbf{N} = \mathbf{{\Lambda}Z}+\mathbf{N}, \label{eq:detected_matrix_pcmb}
\end{align}
where $\mathbf{Y}$ is a $D\times D$ complex-valued matrix, and $\mathbf{N}$ is the $D \times D$ complex-valued additive white Gaussian noise matrix whose elements have zero mean and variance $N_0 = D / SNR$. The channel matrix $\mathbf{H}$ is complex Gaussian with zero mean and unit variance. The total transmitted power is scaled as $D$ in order to make the received Signal-to-Noise Ratio (SNR) $SNR$. Note that in the case of a MIMO system employing PSTBC, the received signal is simply $\mathbf{Y} = \mathbf{HZ}+\mathbf{N}$. With the knowledge of CSIT, the channel matrix $\mathbf{H}$ is now replaced by the diagonal matrix $\mathbf{\Lambda}$ in (\ref{eq:detected_matrix_pcmb}).

Let $\chi$ denote the signal set of the modulation scheme and define $x_{(u,v)}$ as the $(u,v)$th symbol in $\mathbf{X}=[\mathbf{x}_{1}, \ldots, \mathbf{x}_{D}]$ where $u,v \in \{1,\cdots, D \}$. Define the one-to-one mapping from $\mathbf{X}$ to $\mathbf{Z}$ as $\mathbf{Z} = \mathbb{M} \{ \mathbf{X} \}$. Therefore, the ML decoding of (\ref{eq:detected_matrix_pcmb}) is obtained by
\begin{align}
\hat{\mathbf{X}} = \arg\min_{x_{(u,v)} \in \chi, \forall u,v } \| \mathbf{Y}- \mathbf{\Lambda} \mathbb{M} \{ \mathbf{X} \} \| ^2.
\label{eq:ML_decoding_pcmb}
\end{align}

\subsection{BICMB-PC} \label{subsec:BICMB-PC_model}

The structure of BICMB-PC is presented in Fig. \ref{fig:system_model_bicmb-pc}. First, the convolutional encoder of code rate $R_c$, possibly combined with a perforation matrix for a high rate punctured code \cite{Haccoun_PCC}, generates the bit codeword $\mathbf{c}$ from the information bits. A random bit-interleaver is then applied to generate the interleaved bit sequence, which is then modulated by $M$-QAM or $M$-HEX and mapped by Gray encoding. Eventually, $D^2$ consecutive complex-valued scalar symbols are encoded into one PSTBC codeword as (\ref{eq:PSTBC_pcmb}).

Hence, the $k$th PSTBC codeword $\mathbf{Z}_k$ is constructed as
\begin{equation}
\mathbf{Z}_k=\sum_{v=1}^D{\mathrm{diag}(\mathbf{G}\mathbf{x}_{v,k})\mathbf{E}^{v-1}},
\label{eq:PSTBC_bicmb-pc}
\end{equation}
where $\mathbf{x}_{v,k}$ is an $D\times1$ vector whose elements are the $v$th $D$ input modulated scalar symbols to construct the $k$th PSTBC codeword.

The received signal corresponding to the $k$th PSTBC codeword is
\begin{align}
\mathbf{Y}_{k} = \mathbf{{\Lambda}} \mathbf{Z}_k+\mathbf{N}_k \label{eq:detected_matrix_bicmb-pc}
\end{align}
where $\mathbf{Y}_k$, $\mathbf{Z}_k$, and $\mathbf{N}_k$ are the received symbol matrix, the PSTBC codeword, and the noise matrix corresponding to the $k$th PSTBC codeword, respectively.

The location of the coded bit $c_{k'}$ within the PSTBC codeword sequence is denoted as $k' \rightarrow (k, (m,n), j)$, where $k$, $(m,n)$, and $j$ are the index of the PSTBC codewords, the symbol position in $\mathbf{X}_k=[\mathbf{x}_{1,k}, \ldots, \mathbf{x}_{D,k}]$, and the bit position on the label of the scalar symbol $x_{(m,n),k}$, respectively. As defined in Section \ref{subsec:PCMB_model}, $\chi$ denotes the signal set of the modulation scheme. Let $\chi_b^j$ denote a subset of $\chi$ whose labels have $b \in \{0, 1\}$ in the $j$th bit position. By using the location information and the input-output relation in (\ref{eq:detected_matrix_bicmb-pc}), the receiver calculates the ML bit metrics for $c_{k'}=b$ as

\begin{align}
\Gamma^{(m,n),j}(\mathbf{Y}_k, c_{k'}) = \min_{\mathbf{X} \in \eta_{c_{k'}}^{(m,n),j}} \| \mathbf{Y}_k - \mathbf{\Lambda} \mathbb{M} \{ \mathbf{X} \} \|^2, \label{eq:ML_bit_metrics_bicmb-pc}
\end{align}
where $\eta_{c_{k'}}^{(m,n),j}$ is defined as
\begin{align*}
\eta_{b}^{(m,n),j} = \{ \mathbf{X}: x_{(u,v)=(m,n)} \in \chi_{b}^{j}, \, \mathrm{and} \, x_{(u,v) \neq (m,n)} \in \chi \}.
\end{align*}

Finally, the ML decoder, which uses the soft-input Viterbi decoding \cite{Lin_ECC} to find a codeword with the minimum sum weight, makes decisions according to the rule given by \cite{Caire_BICM} as
\begin{align}
\mathbf{\hat{c}} = \arg\min_{\mathbf{c}} \sum_{k'} \Gamma^{(m,n),j}(\mathbf{Y}_k, c_{k'}).
\label{eq:ML_decoding_bicmb-pc}
\end{align} 

%% file: PCMB.tex
\section{PCMB} \label{sec:PCMB}

In this section, the diversity and decoding complexity analyses of GCMB, which is PCMB of dimension $2$, are first investigated in Section \ref{subsec:GCMB_Diversity} and Section \ref{subsec:GCMB_Decoding}, respectively. Then, they are generalized to larger dimensions in Section \ref{subsec:PCMB}. More discussion is provided in Section \ref{subsec:PCMB_Discussions}.

\subsection{Diversity Analysis} \label{subsec:GCMB_Diversity}

For ML decoding, the instantaneous Pairwise Error Probability (PEP) between the transmitted codeword $\mathbf{X}$ and the detected codeword $\hat{\mathbf{X}}$ is represented as
\ifCLASSOPTIONonecolumn
\begin{align}
\mathrm{Pr} \left( \mathbf{X} \rightarrow \hat{\mathbf{X}} \mid \mathbf{H} \right) = \mathrm{Pr} \left( \| \mathbf{Y} - \mathbf{\Lambda} \mathbf{Z} \| ^2 \geq \| \mathbf{Y} - \mathbf{\Lambda} \hat{\mathbf{Z}} \| ^2 \mid \mathbf{H} \right) = \mathrm{Pr} \left( \epsilon \geq \| \mathbf{\Lambda} (\mathbf{Z} - \hat{\mathbf{Z}}) \| ^2 \mid \mathbf{H} \right), \label{eq:PEP_gcmb}
\end{align}
\else
\begin{align}
\mathrm{Pr} \left( \mathbf{X} \rightarrow \hat{\mathbf{X}} \mid \mathbf{H} \right) &= \mathrm{Pr} \left( \| \mathbf{Y} - \mathbf{\Lambda} \mathbf{Z} \| ^2 \geq \| \mathbf{Y} - \mathbf{\Lambda} \hat{\mathbf{Z}} \| ^2 \mid \mathbf{H} \right) \nonumber \\&= \mathrm{Pr} \left( \epsilon \geq \| \mathbf{\Lambda} (\mathbf{Z} - \hat{\mathbf{Z}}) \| ^2 \mid \mathbf{H} \right), \label{eq:PEP_gcmb}
\end{align}
\fi
where $\hat{\mathbf{Z}}=\mathbb{M} \{ \hat{\mathbf{X}} \}$ and $\epsilon = \textrm{Tr} \{ - (\mathbf{Z} - \hat{\mathbf{Z}})^H \mathbf{\Lambda}^H \mathbf{N}- \mathbf{N}^H \mathbf{\Lambda} (\mathbf{Z} - \hat{\mathbf{Z}}) \}$. Since $\epsilon$ is a zero mean Gaussian random variable with variance $2 N_0 \| \mathbf{\Lambda}  (\mathbf{Z} - \hat{\mathbf{Z}}) \| ^2$, (\ref{eq:PEP_gcmb}) is given by the $Q$ function as
\begin{align}
\mathrm{Pr} \left( \mathbf{X} \rightarrow \hat{\mathbf{X}} \mid \mathbf{H} \right) = Q \left( \sqrt{\frac{\| \mathbf{\Lambda} (\mathbf{Z} - \hat{\mathbf{Z}}) \| ^2}{2 N_0}}\right). \label{eq:PEP_gcmb_2}
\end{align}
By using the upper bound on the $Q$ function $Q(x) \leq \frac{1}{2} e^{-x^2/2}$, the average PEP can be upper bounded as
\ifCLASSOPTIONonecolumn
\begin{align}
\mathrm{Pr} \left( \mathbf{X} \rightarrow \hat{\mathbf{X}} \right) = E \left[ \mathrm{Pr} \left( \mathbf{X} \rightarrow
\hat{\mathbf{X}} \mid \mathbf{H} \right) \right] \leq E \left[ \frac{1}{2} \exp \left(- \frac{\| \mathbf{\Lambda} (\mathbf{Z} - \hat{\mathbf{Z}}) \| ^2}{4 N_0}
\right) \right]. \label{eq:PEP_average_gcmb}
\end{align}
\else
\begin{align}
\mathrm{Pr} \left( \mathbf{X} \rightarrow \hat{\mathbf{X}} \right) &= E \left[ \mathrm{Pr} \left( \mathbf{X} \rightarrow
\hat{\mathbf{X}} \mid \mathbf{H} \right) \right] \nonumber \\ &\leq E \left[ \frac{1}{2} \exp \left(- \frac{\| \mathbf{\Lambda} (\mathbf{Z} - \hat{\mathbf{Z}}) \| ^2}{4 N_0}
\right) \right]. \label{eq:PEP_average_gcmb}
\end{align}
\fi
Let $\mathbf{g}^T_u$ with $u \in \{1, 2\}$ denote the $u$th row of $\mathbf{G}$. Then, equation (\ref{eq:PSTBC_pcmb}) can be rewritten as
\begin{align}
\mathbf{Z} =
\left[ \begin{array}{cc}
\mathbf{g}^T_1\mathbf{x}_1 & \mathbf{g}^T_1\mathbf{x}_2 \\
i\mathbf{g}^T_2\mathbf{x}_2 & \mathbf{g}^T_2\mathbf{x}_1
\end{array} \right].
\label{eq:golden_code}
\end{align}
Therefore,
\begin{align}
\mathbf{\Lambda} \mathbf{Z} =
\left[ \begin{array}{cc}
{\lambda}_1\mathbf{g}^T_1\mathbf{x}_1 & {\lambda}_1\mathbf{g}^T_1\mathbf{x}_2 \\
i{\lambda}_2\mathbf{g}^T_2\mathbf{x}_2 & {\lambda}_2\mathbf{g}^T_2\mathbf{x}_1
\end{array} \right].
\label{eq:Lambda_Z}
\end{align}
Then,
\ifCLASSOPTIONtwocolumn
\begin{align}
\| \mathbf{\Lambda} \mathbf{Z} \|^2 =  \textrm{Tr} \left\lbrace \mathbf{Z}^H \mathbf{\Lambda}^H \mathbf{\Lambda Z} \right\rbrace = \sum_{u=1}^{D} {\lambda}_u^2 \sum_{v=1}^{D} | \mathbf{g}^T_u \mathbf{x}_{v} |^2
\label{eq:Lambda_Z_square}
\end{align}
\else
\begin{align}
\| \mathbf{\Lambda} \mathbf{Z} \|^2 =  \textrm{Tr} \left\lbrace \mathbf{Z}^H \mathbf{\Lambda}^H \mathbf{\Lambda Z} \right\rbrace = \sum_{u=1}^{D} {\lambda}_u^2 \sum_{v=1}^{D} | \mathbf{g}^T_u \mathbf{x}_{v} |^2
\label{eq:Lambda_Z_square}
\end{align}
\fi
where $D=2$ for the purposes of (\ref{eq:Lambda_Z_square})-(\ref{eq:PEP_GCMB_final}) in this subsection. As will be discussed later, (\ref{eq:Lambda_Z_square})-(\ref{eq:PEP_GCMB_final}) are actually valid for larger values of $D$ as well. Let $\hat{\mathbf{x}}_1$ and $\hat{\mathbf{x}}_2$ denote the detected symbol vectors. By replacing $\mathbf{x}_1$ and $\mathbf{x}_2$ in (\ref{eq:Lambda_Z_square}) by $\mathbf{x}_1-\hat{\mathbf{x}}_1$ and $\mathbf{x}_2-\hat{\mathbf{x}}_2$, (\ref{eq:PEP_average_gcmb}) is then rewritten as
\begin{align}
\mathrm{Pr} \left( \mathbf{X} \rightarrow \hat{\mathbf{X}} \right) \leq E \left[ \frac{1}{2} \exp \left(- \frac{\sum_{u=1}^{D} {\rho}_u{\lambda}_u^2}{4 N_0}
\right) \right], \label{eq:PEP_average_gcmb_2}
\end{align}
where
\begin{align}
{\rho}_u=\sum_{v=1}^{D} | \mathbf{g}^T_u (\mathbf{x}_{v} - \hat{\mathbf{x}}_{v})|^2.
\label{eq:Weight_GCMB}
\end{align}

The upper bound in (\ref{eq:PEP_average_gcmb_2}) can be further bounded by employing a theorem from \cite{Park_UP_MPDF} which is given below.
\begin{theorem}
Consider the largest $S \leq \min(N_t, N_r)$ eigenvalues $\mu_s$ of the uncorrelated central $N_r \times N_t$ Wishart matrix that are sorted in decreasing order, and a weight vector $\boldsymbol{\rho} = [\rho_1, \cdots, \rho_S]^T$ with non-negative real elements. In the high SNR regime, an upper bound for the expression $E [ \exp (-\gamma \sum_{s=1}^S \rho_s \mu_s ) ]$, which is used in the diversity analysis of a number of MIMO systems, is
\begin{align*}
E\left[ \exp \left( - \gamma \sum\limits_{s=1}^S \rho_s \mu_s \right) \right] \leq \zeta \left( \rho_{min} \gamma \right)^{-(N_r-\delta+1)(N_t-\delta+1)},
\end{align*}
where $\gamma$ is SNR, $\zeta$ is a constant, $\rho_{min} = \min_{\rho_i \neq 0} {\{ \rho_i \}}_{i=1}^{S}$, and $\delta$ is the index to the first non-zero element in the weight vector.
\label{theorem:E_PEP}
\end{theorem}
\begin{IEEEproof}
See \cite{Park_UP_MPDF}.
\end{IEEEproof}

Based on the aforementioned theorem, full diversity is achieved if and only if $\delta=1$, which is equivalent to $\rho_1 > 0$. Note that ${\rho}_1 > 0$ in (\ref{eq:Weight_GCMB}) because all elements in $\mathbf{g}_1^T$ are nonzero \cite{Belfiore_GC}, and therefore $\delta=1$. By applying the Theorem 
to (\ref{eq:PEP_average_gcmb_2}), an upper bound of PEP is
\begin{align}
\mathrm{Pr} \left( \mathbf{X} \rightarrow \hat{\mathbf{X}} \right) &\leq \zeta \left( \frac{\min\{\rho_u \}_{u=1}^{D}}{4 D} SNR \right)^{-N_rN_t}.
\label{eq:PEP_GCMB_final}
\end{align}
Since $N_t=N_r=D=2$ in this case, GCMB achieves the full diversity order of $4$.

\subsection{Decoding} \label{subsec:GCMB_Decoding}
Equation (\ref{eq:Lambda_Z}) shows that each element of $\mathbf{\Lambda} \mathbf{Z}$ is only related to $\mathbf{x}_1$ or $\mathbf{x}_2$. Consequently, the elements of $\mathbf{\Lambda} \mathbf{Z}$ can be divided into two groups, and the first and second groups contain elements related to $\mathbf{x}_1$ and $\mathbf{x}_2$, respectively. The input-output relation in (\ref{eq:detected_matrix_pcmb}) then is decomposed into two equations as
\begin{align}
\begin{split}
&\breve{\mathbf{y}}_1 = \left[
\begin{array}{c}
y_{(1,1)} \\
y_{(2,2)}
\end{array} \right]  = \left[
\begin{array}{c}
{\lambda}_1\mathbf{g}^T_1\mathbf{x}_1 \\
{\lambda}_2\mathbf{g}^T_2\mathbf{x}_1
\end{array} \right] + \left[
\begin{array}{c}
n_{(1,1)} \\
n_{(2,2)}
\end{array} \right], \\
&\breve{\mathbf{y}}_2 = \left[
\begin{array}{c}
y_{(1,2)} \\
y_{(2,1)}
\end{array} \right]  = \left[
\begin{array}{c}
{\lambda}_1\mathbf{g}^T_1\mathbf{x}_2 \\
i{\lambda}_2\mathbf{g}^T_2\mathbf{x}_2
\end{array} \right] + \left[
\begin{array}{c}
n_{(1,2)} \\
n_{(2,1)}
\end{array} \right],
\end{split} \label{eq:deteced_symbol_decomposed_gcmb}
\end{align}
Let $\breve{\mathbf{n}}_1=[n_{(1,1)}, n_{(2,2)}]^T$ and $\breve{\mathbf{n}}_2=[n_{(1,2)}, n_{(2,1)}]^T$, then (\ref{eq:deteced_symbol_decomposed_gcmb}) can be further rewritten as
\begin{align}
\begin{split}
&\breve{\mathbf{y}}_1 = \mathbf{\Lambda G x_1} + \breve{\mathbf{n}}_1, \\
&\breve{\mathbf{y}}_2 = \mathbf{\Phi \Lambda G x_2} + \breve{\mathbf{n}}_2,
\end{split} \label{eq:deteced_symbol_decomposed_gcmb_2}
\end{align}
where
\begin{align*}
\mathbf{\Phi} = \left[\
\begin{array}{cc}
1 & 0 \\
0 & i
\end{array} \right].
\end{align*}

The input-output relation of (\ref{eq:deteced_symbol_decomposed_gcmb_2}) implies that the threaded structure of the codeword in (\ref{eq:PSTBC_pcmb}) is now separated with the knowledge of CSIT, and therefore $\mathbf{x}_1$ and $\mathbf{x}_2$ can be decoded independently.

By using the QR decomposition of $\mathbf{\Lambda G}=\mathbf{QR}$, where $\mathbf{R}$ is an upper triangular matrix, and the matrix $\mathbf{Q}$ is unitary, (\ref{eq:deteced_symbol_decomposed_gcmb_2}) is rewritten as
\begin{align}
\begin{split}
&\tilde{\mathbf{y}}_1 = \mathbf{Q}^H \breve{\mathbf{y}}_1 = \mathbf{R}\mathbf{x}_1 + \mathbf{Q}^H \breve{\mathbf{n}}_1 = \mathbf{R}\mathbf{x}_1 + \tilde{\mathbf{n}}_1, \\
&\tilde{\mathbf{y}}_2 = \mathbf{Q}^H \mathbf{\Phi}^H \breve{\mathbf{y}}_2 = \mathbf{R}\mathbf{x}_2 + \mathbf{Q}^H \mathbf{\Phi}^H \breve{\mathbf{n}}_2 = \mathbf{R}\mathbf{x}_2 + \tilde{\mathbf{n}}_2.
\end{split} \label{eq:deteced_symbol_decomposed_gcmb_3}
\end{align}

Indeed, each relation of (\ref{eq:deteced_symbol_decomposed_gcmb_3}) has the same form as FPMB presented in \cite{Park_CPB}, \cite{Park_MB_CP}, \cite{Park_CPMB}, which is the state-of-the-art full-diversity full-multiplexing SVD-based uncoded technique.
FPMB is the special case of Constellation Precoded Multiple Beamforming (CPMB) whose system model is presented in Fig. \ref{fig:system_model_cpmb}, when the number of precoded symbol streams equals to the number of employed subchannels. In Fig. \ref{fig:system_model_cpmb}, $\mathbf{\Theta}_P$ is the constellation precoding matrix to precode $P$ symbol streams, and $\mathbf{T}$ is a permutation matrix to select precoded subchannels. In \cite{Azzam_SD_NLR}, \cite{Azzam_SD_RLR}, a reduced complexity SD is introduced. The technique takes advantage of a special real lattice representation, which introduces orthogonality between the real and imaginary parts of each symbol, thus enables employing rounding (or quantization) for the last two layers of the SD. When the dimension is $2\times2$, it achieves ML performance with the worst-case decoding complexity of $\mathcal{O}(M)$. This technique can be employed to decode both GCMB and $2\times2$ FPMB, since their input-output relations can be written in the same form as (\ref{eq:deteced_symbol_decomposed_gcmb_3}).

Furthermore, lower decoding complexity can be achieved for GCMB because of the special property of the $\mathbf{G}$ matrix. The $\mathbf{G}$ matrix for dimension $2$ is given by \cite{Belfiore_GC} as
\begin{align*}
\mathbf{G}=\frac{1}{\sqrt{5}} \left[
\begin{array}{cc}
1+i\beta & {\alpha}-i \\
1+i\alpha & {\beta}-i
\end{array} \right],
\end{align*}
with $\alpha={1+\sqrt{5} \over 2}$ and $\beta={1-\sqrt{5} \over 2}$. Let $\mathbf{f}_v$ denote the $v$th column of
\begin{align}
\mathbf{\Lambda G} = \frac{1}{\sqrt{5}} \left[
\begin{array}{cc}
\lambda_1(1+i\beta) & \lambda_1({\alpha}-i) \\
\lambda_2(1+i\alpha) & \lambda_2({\beta}-i)
\end{array} \right],
\label{eq:Lambda_G}
\end{align}
where $v \in \{1, 2\}$. The nonzero elements of the diagonal matrix $\mathbf{R}$ are calculated as
\begin{equation}
\begin{split}
&r_{(1,1)}=\| \mathbf{f}_1 \|, \\
&r_{(1,2)}=\frac{\mathbf{f}_2^H\mathbf{f}_1}{\| \mathbf{f}_1 \|}=\frac{({\alpha}-{\beta})({\lambda}_1^2-{\lambda}_2^2)}{5 \| \mathbf{f}_1 \|}, \\
&r_{(2,2)}=\left\| \mathbf{f}_2-\frac{ \mathbf{f}_1^H\mathbf{f}_2} { \| \mathbf{f}_1 \| ^2 } \mathbf{f}_1 \right\|.
\label{eq:R_elements}
\end{split}
\end{equation}
Note that $\mathbf{R}$ is a complex-valued matrix in general when the QR decomposition is applied to a complex-valued matrix. However, based on (\ref{eq:R_elements}), the $\mathbf{R}$ matrix is real-valued for GCMB, which is due to the special property of the  $\mathbf{G}$ matrix. Hence, the real and imaginary parts of (\ref{eq:deteced_symbol_decomposed_gcmb_3}) can be decoded separately. Consequently, (\ref{eq:deteced_symbol_decomposed_gcmb_3}) can be decomposed further as
\begin{align}
\begin{split}
&\Re \{\tilde{\mathbf{y}}_u\} = \mathbf{R}\Re \{\mathbf{x}_u\} + \Re \{\tilde{\mathbf{n}}_u\}, \\
&\Im \{\tilde{\mathbf{y}}_u\} = \mathbf{R}\Im \{\mathbf{x}_u\} + \Im \{\tilde{\mathbf{n}}_u\}, \\
\end{split} \label{eq:deteced_symbol_decomposed_gcmb_4}
\end{align}
with $u \in \{1,2\}$. To decode each part of (\ref{eq:deteced_symbol_decomposed_gcmb_4}), a two-level real-valued SD can be employed plus applying the rounding procedure for the last layer. As a result, the worst-case decoding complexity of GCMB is $\mathcal{O}(\sqrt{M})$.

Previously, the ML decoding of GC was shown to have the worst-case complexity of $\mathcal{O}(M^{2.5})$ \cite{Sinnokrot_FMLD_GC}, \cite{Sinnokrot_GC_FD}, \cite{Sinnokrot_STBC_LMLDC}. However, the above analysis proves that this complexity can be reduced substantially to only $\mathcal{O}(\sqrt{M})$ by applying GCMB when CSIT is known. Furthermore, the complexity of GCMB is lower than FPMB as well. The worst-case decoding complexity of $2\times2$ FPMB with the decoding technique presented in \cite{Azzam_SD_NLR}, \cite{Azzam_SD_RLR} is $\mathcal{O}(M)$ as mentioned above.

\subsection{PCMB} \label{subsec:PCMB}
For PCMB of dimension $D \in \{3,4,6\}$, it can be proved that they all achieve the full diversity order of $D^2$, which is generalized from (\ref{eq:Lambda_Z_square})-(\ref{eq:PEP_GCMB_final}) because they are still valid for larger $D$.

For the decoding of PCMB in dimension $D \in \{3,4,6\}$, similarly to GCMB, the elements of $\mathbf{\Lambda} \mathbf{Z}$ are related to only one of the $\mathbf{x}_v$, thus can be divided into $D$ groups, where the $v$th group contains elements related to $\mathbf{x}_v$. The received signal is then divided into $D$ parts, which can be represented as
\begin{align}
\mathbf{y}_v = \mathbf{\Phi}_v \mathbf{\Lambda G} \mathbf{x}_v + \mathbf{n}_v,
\label{eq:deteced_symbol_decomposed_pcmb}
\end{align}
where $\mathbf{\Phi}_v=\textrm{diag}(\phi_{v,1}, \cdots, \phi_{v,D})$ is a diagonal unitary matrix whose elements satisfy
\begin{align*}
{\phi}_{v,k} = \left\lbrace
\begin{array}{cc}
1, & 1 \leq k \leq D+1-v, \\
g, & D+2-v \leq k \leq D.
\end{array} \right.
\end{align*}
By using the QR decomposition of $\mathbf{\Lambda G}=\mathbf{Q} \mathbf{R}$, and moving $\mathbf{\Phi}_v\mathbf{Q}$ to the left hand, (\ref{eq:deteced_symbol_decomposed_pcmb}) is rewritten as
\begin{align}
\tilde{\mathbf{y}}_v = \mathbf{Q}^H \mathbf{\Phi}_v^H \mathbf{y}_v = \mathbf{R}\mathbf{x}_v + \mathbf{Q}^H \mathbf{\Phi}_v^H \mathbf{n}_v = \mathbf{R}\mathbf{x}_v + \tilde{\mathbf{n}}_v.
\label{eq:deteced_symbol_decomposed_pcmb_2}
\end{align}

For the dimension of $4$, the $\mathbf{R}$ matrix in (\ref{eq:deteced_symbol_decomposed_pcmb_2}) is real-valued, which can be proved in a similar way to GCMB in Section \ref{subsec:GCMB_Diversity}. See the Appendix for the proof. Consequently, the real part and the imaginary part of $\mathbf{x}_v$ can be decoded separately as (\ref{eq:deteced_symbol_decomposed_gcmb_4}) with $u \in \{1, \ldots, 4\}$. Real-valued SD with the last layer rounded can be employed, and the worst-case decoding complexity of PCMB is then $\mathcal{O}(M^{1.5})$. Regarding a MIMO system employing PSTBC, the worst-case decoding complexity is $\mathcal{O}(M^{13.5})$ by using the technique presented in \cite{Sinnokrot_STBC_LMLDC}. For FPMB, ML decoding can be achieved by using SD based on the real lattice representation in \cite{Azzam_SD_NLR}, \cite{Azzam_SD_RLR}, plus quantization of the last two layers, and the worst-case complexity is then $\mathcal{O}(M^{3})$.

Unfortunately, for the $D=3$ ($6$) dimension case, the $\mathbf{R}$ matrix is complex-valued. Therefore, the real and the imaginary parts of $\mathbf{x}_v$ cannot be decoded separately, unlike the case of $D=2,4$. Moreover, since the M-HEX modulation is employed, which cannot be separated as two independent one-dimensional modulations as in M-QAM, a complex-valued SD, instead of a real-valued SD, with an efficient implementation of a slicer \cite{Sinnokrot_STBC_LMLDC} is applied. The worst-case decoding complexity of PCMB is then $\mathcal{O}(M^{2})$ ($\mathcal{O}(M^{5})$). In the case of a MIMO system employing PSTBC, the worst-case decoding complexity is $\mathcal{O}(M^{8})$ ($\mathcal{O}(M^{35})$) \cite{Sinnokrot_STBC_LMLDC}. For FPMB, the worst-case decoding complexity is $\mathcal{O}(M^{2})$ ($\mathcal{O}(M^{5})$), which is similar to PCMB.


\subsection{Discussion} \label{subsec:PCMB_Discussions}

\ifCLASSOPTIONonecolumn
\begin{table}[!m]
\renewcommand{\arraystretch}{1.3}
\caption{Worst-Case Complexity}
\centering
\label{table:complexity}
\begin{tabular}{|c|c|c|c|c|}
\hline
& $D=2$ & $D=3$ & $D=4$ & $D=6$ \\
\hline
PC & $\mathcal{O}(M^{2.5})$ & $\mathcal{O}(M^{8})$ & $\mathcal{O}(M^{13.5})$ & $\mathcal{O}(M^{35})$ \\
\hline
FPMB (BICMB-FP) & $\mathcal{O}(M)$ & $\mathcal{O}(M^{2})$ & $\mathcal{O}(M^{3})$ & $\mathcal{O}(M^{5})$ \\
\hline
PCMB (BICMB-PC) & $\mathcal{O}(\sqrt{M})$ & $\mathcal{O}(M^{2})$ & $\mathcal{O}(M^{1.5})$ & $\mathcal{O}(M^{5})$ \\
\hline
\end{tabular}
\end{table}
\else
\begin{table}[!t]
\renewcommand{\arraystretch}{1.3}
\caption{Worst-Case Complexity}
\centering
\label{table:complexity}
\begin{tabular}{|c|c|c|c|c|}
\hline
& $D=2$ & $D=3$ & $D=4$ & $D=6$ \\
\hline
PC & $\mathcal{O}(M^{2.5})$ & $\mathcal{O}(M^{8})$ & $\mathcal{O}(M^{13.5})$ & $\mathcal{O}(M^{35})$ \\
\hline
FPMB & $\mathcal{O}(M)$ & $\mathcal{O}(M^{2})$ & $\mathcal{O}(M^{3})$ & $\mathcal{O}(M^{5})$ \\
\hline
PCMB & $\mathcal{O}(\sqrt{M})$ & $\mathcal{O}(M^{2})$ & $\mathcal{O}(M^{1.5})$ & $\mathcal{O}(M^{5})$ \\
\hline
BICMB-FP & $\mathcal{O}(M)$ & $\mathcal{O}(M^{2})$ & $\mathcal{O}(M^{3})$ & $\mathcal{O}(M^{5})$ \\
\hline
BICMB-PC & $\mathcal{O}(\sqrt{M})$ & $\mathcal{O}(M^{2})$ & $\mathcal{O}(M^{1.5})$ & $\mathcal{O}(M^{5})$ \\
\hline
\end{tabular}
\end{table}
\fi

Table \ref{table:complexity} summarizes the worst-case complexity of a MIMO system employing PSTBC which is denoted by PC, FPMB, and PCMB for different dimensions to decode one received symbol vector.

As shown in Table \ref{table:complexity}, the decoding complexity of PCMB is substantially lower than PC. Actually, the problem of high decoding complexity results from the threaded structure of PSTBCs in (\ref{eq:PSTBC_pcmb}). With the knowledge of CSIT, PCMB successfully separates the threaded structure of PSTBCs at the receiver, and thereby reduces the dimension of the decoding problem from $D^2$ to $D$, as (\ref{eq:deteced_symbol_decomposed_gcmb_3}) and (\ref{eq:deteced_symbol_decomposed_pcmb_2}), which mainly results in the complexity advantage of PCMB over PC. This result provides a new prospect of CSIT, which could also be applied to reduce the decoding complexity of a MIMO system, since it is mostly applied to either increase the throughput or to enhance performance previously.

Nevertheless, there are always tradeoffs among throughput, reliability, and complexity for MIMO systems in general \cite{Zheng_DM}, \cite{Jafarkhani_STC}. In fact, the nonvanishing constant minimum determinant of PSTBCs, which offers high coding gain, is also derived from the threaded structure. As a result, this property is no longer valid for PCMB, which sacrifices the coding gain. The coding gain loss is hard to quantify, but simulation results in Section \ref{subsec:Results_PCMB} show that only negligible or modest loss is caused. Other than the nonvanishing constant minimum determinant, other good properties of PSTBCs, i.e., full rate, full diversity, uniform average transmitted energy per antenna, and good shaping, are still valid for PCMB.

In fact, since PSTBCs belong to the class of Threaded Algebraic Space-Time (TAST) codes \cite{Gamal_USTC}, and CSIT results in separating the threaded structure at the receiver, the same idea can be applied to reduce the decoding complexity of general TAST codes as well. However, PCMB of dimensions $2$ and $4$ have a further advantage in terms of decoding complexity due to the real-valued upper triangular $\mathbf{R}$ matrices in (\ref{eq:deteced_symbol_decomposed_gcmb_3}) and (\ref{eq:deteced_symbol_decomposed_pcmb_2}), which leads to the separate decoding of the real and the imaginary parts as (\ref{eq:deteced_symbol_decomposed_gcmb_4}). In other words, the $D$-dimensional complex-valued decoding problem can be further decomposed into two $D$-dimensional real-valued decoding problems as (\ref{eq:deteced_symbol_decomposed_gcmb_4}). This advantage is related to the special property of the generation matrices of PSTBCs in these two dimensions, and thereby is not valid for general TAST codes.

Note that full diversity of PCMB results from the fact that all elements of $\mathbf{g}_1^T$ are nonzero, the complexity advantages of PCMB over PC is due to the knowledge of CSIT and the threaded structure of PSTBCs, and the additional complexity reduction of PCMB in dimension $2$ and $4$ is caused by the real-valued $\mathbf{R}$ matrices. Since they are not related to some good properties of PSTBCs such as uniform average transmitted energy per antenna and good shaping, there may exist other space-time block codes which can achieve the same advantages of PCMB.

CSIT is assumed to be known for both PCMB and FPMB, and both of them achieve full diversity and full multiplexing. In dimensions $3$ and $6$, the decoding complexity of PCMB is similar to FPMB. However, PCMB has significant decoding complexity advantage in dimensions $2$ and $4$, due to their real-valued $\mathbf{R}$ matrices, over FPMB, whose $\mathbf{R}$ matrices are complex-valued. Similarly, since FPMB is designed to achieve high array gains \cite{Park_CPB}, while PCMB does not concentrate on this aspect, tradeoffs between complexity and array gain might exist. The array gain is hard to quantify, but simulation results in Section \ref{subsec:Results_PCMB} show that only negligible or modest loss is caused by PCMB compared to FPMB. 

%% file: BICMB-PC.tex
\section{BICMB-PC} \label{sec:BICMB-PC}
In this section, the diversity and decoding complexity analyses of BICMB-GC, which is BICMB-PC of dimension $2$, are first carried out in Section \ref{subsec:BICMB-GC_Diversity} and Section \ref{subsec:BICMB-GC_Decoding}, respectively. Then, they are generalized to larger dimensions in Section \ref{subsec:BICMB-PC}. More discussion is provided in \ref{subsec:BICMB-PC_Discussions}.

\subsection{Diversity Analysis} \label{subsec:BICMB-GC_Diversity}
Based on the bit metrics in (\ref{eq:ML_bit_metrics_bicmb-pc}), the instantaneous PEP of BICMB-GC between the transmitted bit codeword $\mathbf{c}$ and the decoded bit codeword $\mathbf{\hat{c}}$ is
\begin{align}
\mathrm{Pr} \left( \mathbf{c} \rightarrow \hat{\mathbf{c}} \mid \mathbf{H} \right) = \mathrm{Pr} \left( \sum_{k'} \min_{\mathbf{X} \in \eta_{c_{k'}}^j} \| \mathbf{Y}_k - \mathbf{\Lambda} \mathbb{M} \{ \mathbf{X} \} \|^2 \geq \right. \ifCLASSOPTIONtwocolumn \nonumber \\ \fi \left. \sum_{k'} \min_{\mathbf{X} \in \eta_{\hat{c}_{k'}}^j} \| \mathbf{Y}_k - \mathbf{\Lambda} \mathbb{M} \{ \mathbf{X} \}\|^2 \mid \mathbf{H} \right), \label{eq:PEP_original_bicmb-gc}
\end{align}
where $c_{k'}$ and $\hat{c}_{k'}$ are the coded bit of $\mathbf{c}$ and $\mathbf{\hat{c}}$, respectively. Let $d_H$ denote the Hamming distance between $\mathbf{c}$ and $\mathbf{\hat{c}}$. 
Since the bit metrics corresponding to the same coded bits between the pairwise errors are the same, (\ref{eq:PEP_original_bicmb-gc}) is rewritten as
\begin{align}
\mathrm{Pr} \left(\mathbf{c} \rightarrow \hat{\mathbf{c}} \mid \mathbf{H}\right) = \mathrm{Pr} \left( \sum_{k', d_H} \min_{\mathbf{X} \in \eta_{c_{k'}}^j} \| \mathbf{Y}_k - \mathbf{\Lambda} \mathbb{M} \{ \mathbf{X} \} \|^2 \geq \right. \ifCLASSOPTIONtwocolumn \nonumber \\ \fi \left. \sum_{k', d_H} \min_{\mathbf{X} \in \eta_{\hat{c}_{k'}}^j} \| \mathbf{Y}_k - \mathbf{\Lambda} \mathbb{M} \{ \mathbf{X} \} \|^2 \mid \mathbf{H} \right),
\label{eq:PEP_for_different_codedbits_bicmb-gc}
\end{align}
where $\sum_{k', d_H}$ stands for the summation of the $d_H$ values corresponding to the different coded bits between the bit codewords.

Define $\tilde{\mathbf{X}}_k$ and $\hat{\mathbf{X}}_k$ as
\begin{equation}
\begin{split}
\tilde{\mathbf{X}}_k = \arg\min_{\mathbf{X} \in \eta_{c_{k'}}^{j}} \| \mathbf{Y}_k - \mathbf{\Lambda} \mathbb{M} \{ \mathbf{X} \} \|^2, \\
\hat{\mathbf{X}}_k = \arg\min_{\mathbf{X} \in \eta_{\bar{c}_{k'}}^{j}} \| \mathbf{Y}_k - \mathbf{\Lambda} \mathbb{M} \{ \mathbf{X} \} \|^2.
\end{split}
\label{eq:arg_min}
\end{equation}
It is easily found that $\tilde{\mathbf{X}}_k$ is different from $\hat{\mathbf{X}}_k$ since the sets that $x_{(m,n)}$ belong to are disjoint, as can be seen from the definition of $\eta_{c_{k'}}^{j}$. In the same manner, it is clear that $\mathbf{X}_k$ is different from $\hat{\mathbf{X}}_k$. With $\tilde{\mathbf{Z}}_k=\mathbb{M} \{ \tilde{\mathbf{X}}_k \}$ and $\hat{\mathbf{Z}}_k=\mathbb{M} \{ \hat{\mathbf{X}}_k \}$, (\ref{eq:PEP_for_different_codedbits_bicmb-gc}) is rewritten as
\begin{align}
\mathrm{Pr} \left( \mathbf{c} \rightarrow \mathbf{\hat{c}} \mid \mathbf{H} \right) = \mathrm{Pr} \left( \sum_{k', d_H} \| \mathbf{Y}_k - \mathbf{\Lambda} \tilde{\mathbf{Z}}_k \|^2 \geq \right. \ifCLASSOPTIONtwocolumn \nonumber \\ \fi \left. \sum_{k', d_H} \| \mathbf{Y}_k - \mathbf{\Lambda} \hat{\mathbf{Z}}_k \|^2 \right).
\label{eq:alt_expression_PEP_diffbits_bicmb-gc}
\end{align}
Based on the fact that $\| \mathbf{Y}_k - \mathbf{{\Lambda}} \mathbf{Z}_k \|^2 \geq  \| \mathbf{Y}_k - \mathbf{{\Lambda}} \mathbf{\tilde{Z}}_k \|^2$, and the relation in (\ref{eq:detected_matrix_bicmb-pc}), equation (\ref{eq:alt_expression_PEP_diffbits_bicmb-gc}) is upper bounded by
\ifCLASSOPTIONonecolumn
\begin{align}
\mathrm{Pr} (\mathbf{c} \rightarrow \mathbf{\hat{c}} \mid \mathbf{H}) \leq \mathrm{Pr} \left( \sum_{k', d_H} \| \mathbf{Y}_k - \mathbf{\Lambda} {\mathbf{Z}}_k \|^2 \geq \right. \ifCLASSOPTIONtwocolumn \nonumber \\ \fi \left. \sum_{k', d_H} \| \mathbf{Y}_k - \mathbf{\Lambda} \hat{\mathbf{Z}}_k \|^2 \right) \ifCLASSOPTIONtwocolumn \nonumber \\ \fi = \mathrm{Pr} \left( \epsilon \geq \sum_{k', d_H} \| \mathbf{\Lambda} (\mathbf{Z}_k - \mathbf{\hat{Z}}_k) \|^2 \right), \label{eq:PEP_upperbounded_bicmb-gc}
\end{align}
\else
\begin{align}
\mathrm{Pr} (\mathbf{c} \rightarrow \mathbf{\hat{c}} \mid \mathbf{H}) & \leq \mathrm{Pr} \left( \sum_{k', d_H} \| \mathbf{Y}_k - \mathbf{\Lambda} {\mathbf{Z}}_k \|^2 \geq \right. \ifCLASSOPTIONtwocolumn \nonumber \\ \fi & \qquad \quad \, \left. \sum_{k', d_H} \| \mathbf{Y}_k - \mathbf{\Lambda} \hat{\mathbf{Z}}_k \|^2 \right) \ifCLASSOPTIONtwocolumn \nonumber \\ \fi & = \mathrm{Pr} \left( \epsilon \geq \sum_{k', d_H} \| \mathbf{\Lambda} (\mathbf{Z}_k - \mathbf{\hat{Z}}_k) \|^2 \right), \label{eq:PEP_upperbounded_bicmb-gc}
\end{align}
\fi
where $\epsilon = \sum_{k', d_H} \mathrm{Tr} [ - (\mathbf{Z}_k - \mathbf{\hat{Z}_k})^H \mathbf{\Lambda}^H \mathbf{N}_k- \mathbf{N}_k^H \mathbf{\Lambda} (\mathbf{Z}_k - \mathbf{\hat{Z}}_k) ]$. Since $\epsilon$ is a zero-mean Gaussian random variable with variance $2 N_0 \sum_{k', d_H} \| \mathbf{\Lambda}  (\mathbf{Z}_k - \mathbf{\hat{Z}}_k) \| ^2$, the average PEP can be upper bounded in a similar fashion to (\ref{eq:PEP_gcmb_2}) and (\ref{eq:PEP_average_gcmb}) as
\ifCLASSOPTIONonecolumn
\begin{align}
\mathrm{Pr} \left( \mathbf{c} \rightarrow \mathbf{\hat{c}} \right) = E \left[ \mathrm{Pr} \left( \mathbf{c} \rightarrow \mathbf{\hat{c}} \mid \mathbf{H} \right) \right]
\leq E \left[ \frac{1}{2} \exp \left(- \frac{\sum_{k', d_H} \| \mathbf{\Lambda} (\mathbf{Z}_k - \mathbf{\hat{Z}}_k) \| ^2}{4 N_0} \right) \right].
\label{eq:PEP_average_bicmb-gc}
\end{align}
\else
\begin{align}
\mathrm{Pr} \left( \mathbf{c} \rightarrow \mathbf{\hat{c}} \right) &= E \left[ \mathrm{Pr} \left( \mathbf{c} \rightarrow \mathbf{\hat{c}} \mid \mathbf{H} \right) \right] \nonumber \\
&\leq E \left[ \frac{1}{2} \exp \left(- \frac{\sum_{k', d_H} \| \mathbf{\Lambda} (\mathbf{Z}_k - \mathbf{\hat{Z}}_k) \| ^2}{4 N_0} \right) \right].
\label{eq:PEP_average_bicmb-gc}
\end{align}
\fi
According to (\ref{eq:Lambda_Z_square}), (\ref{eq:PEP_average_bicmb-gc}) is rewritten as
\begin{align}
\mathrm{Pr} \left( \mathbf{c} \rightarrow \mathbf{\hat{c}} \right) \leq E \left[ \frac{1}{2} \exp \left(- \frac{ \sum_{u=1}^{D}\lambda_u^2 \sum_{k', d_H} \rho_{u,k}} {4 N_0} \right) \right],
\label{eq:PEP_average_bicmb-gc_2}
\end{align}
where
\begin{align}
{\rho}_{u,k}=\sum_{v=1}^{D}| \mathbf{g}^T_u(\mathbf{x}_{v,k}-\hat{\mathbf{x}}_{v,k}) |^2, \label{eq:Weight_BICMB-GC}
\end{align}
and $D=2$ for the purposes of (\ref{eq:PEP_average_bicmb-gc_2})-(\ref{eq:PEP_PSB_final_bicmb-gc}) in this subsection. As will be discussed later, (\ref{eq:PEP_average_bicmb-gc_2})-(\ref{eq:PEP_PSB_final_bicmb-gc}) are actually valid for larger values of $D$ as well.

Applying the theorem presented in Section \ref{subsec:GCMB_Diversity} to (\ref{eq:PEP_average_bicmb-gc_2}), $\delta=1$ because $\rho_{1,k} > 0$ in (\ref{eq:Weight_BICMB-GC}). Therefore, an upper bound of PEP is
\begin{align}
\mathrm{Pr} \left( \mathbf{c} \rightarrow \mathbf{\hat{c}} \right) &\leq \zeta \left( \frac{\min\{\sum_{k', d_H} \rho_{u,k},\}_{u=1}^{D}}{4 D} SNR \right)^{-N_rN_t}.
\label{eq:PEP_PSB_final_bicmb-gc}
\end{align}
Since $N_t=N_r=D=2$ in this case, BICMB-GC achieves the full diversity order of $4$.

\subsection{Decoding} \label{subsec:BICMB-GC_Decoding}
Similarly to (\ref{eq:Lambda_Z}), each element of $\mathbf{\Lambda} \mathbf{Z}_k$ for BICMB-GC in (\ref{eq:detected_matrix_bicmb-pc}) is related to only $\mathbf{x}_{1,k}$ or $\mathbf{x}_{2,k}$. Consequently, the elements of $\mathbf{\Lambda} \mathbf{Z}_k$ can be divided into two groups, and the first and second groups contain elements related to $\mathbf{x}_{1,k}$ and $\mathbf{x}_{2,k}$, respectively. The input-output relation in (\ref{eq:detected_matrix_bicmb-pc}) is then decomposed into two equations similarly to (\ref{eq:deteced_symbol_decomposed_gcmb_2}) as
\begin{align}
\begin{split}
&\breve{\mathbf{y}}_{1,k} = \mathbf{\Lambda G} \mathbf{x}_{1,k} + \breve{\mathbf{n}}_{1,k}, \\
&\breve{\mathbf{y}}_{2,k} = \mathbf{\Phi \Lambda G} \mathbf{x}_{2,k} + \breve{\mathbf{n}}_{2,k},
\end{split} \label{eq:deteced_symbol_decomposed_bicmb-gc}
\end{align}
where $\breve{\mathbf{y}}_{1,k}=[Y_{(1,1),k}, Y_{(2,2),k}]^T$, $\breve{\mathbf{y}}_{2,k}=[Y_{(1,2),k}, Y_{(2,1),k}]^T$, $\breve{\mathbf{n}}_{1,k}=[N_{(1,1),k}, N_{(2,2),k}]^T$, and $\breve{\mathbf{n}}_{2,k}=[N_{(1,2),k},$ $N_{(2,1),k}]^T$, with  $Y_{(m,n),k}$ and $N_{(m,n),k}$ denoting the $(m,n)$th element of $\mathbf{Y}_k$ and $\mathbf{N}_k$, respectively.

By using the QR decomposition of $\mathbf{\Lambda G}=\mathbf{Q} \mathbf{R}$, 
(\ref{eq:deteced_symbol_decomposed_bicmb-gc}) is rewritten as
\begin{align}
\begin{split}
&\tilde{\mathbf{y}}_{1,k} = \mathbf{Q}^H \breve{\mathbf{y}}_{1,k} = \mathbf{R}\mathbf{x}_{1,k} + \mathbf{Q}^H \breve{\mathbf{n}}_{1,k} = \mathbf{R}\mathbf{x}_{1,k} + \tilde{\mathbf{n}}_{1,k}, \\
&\tilde{\mathbf{y}}_{2,k} = \mathbf{Q}^H \mathbf{\Phi}^H \breve{\mathbf{y}}_{2,k} = \mathbf{R}\mathbf{x}_{2,k} + \mathbf{Q}^H \mathbf{\Phi}^H \breve{\mathbf{n}}_{2,k} = \mathbf{R}\mathbf{x}_{2,k} + \tilde{\mathbf{n}}_{2,k}.
\end{split} \label{eq:deteced_symbol_decomposed_bicmb-gc_2}
\end{align}
Then the ML bit metrics in (\ref{eq:ML_bit_metrics_bicmb-pc}) can be simplified as
\begin{align}
\Gamma^{(m,n),j}(\mathbf{Y}_k, c_{k'}) = \min_{\mathbf{x} \in \xi_{c_{k'}}^{n,j}} \| \tilde{\mathbf{y}}_{m,k} - \mathbf{R} \mathbf{x} \|^2, \label{eq:ML_bit_metrics_bicmb-pc_2}
\end{align}
where $\xi_{c_{k'}}^{n,j}$ is a subset of $\chi^D$, defined as
\begin{align*}
\xi_{b}^{n,j} = \{ \mathbf{x} = [x_1 \, \cdots \, x_D ]^T : x_{d=n} \in \chi_{b}^{j} \mathrm{ \ and \  } x_{d \neq n} \in \chi \}.
\end{align*}

Indeed, the simplified ML bit metrics (\ref{eq:ML_bit_metrics_bicmb-pc_2}) have the same form as BICMB-FP presented in \cite{Park_BICMB_CP}, \cite{Park_MB_CP}, \cite{Park_CPMB}, which is the state-of-the-art full-diversity full-multiplexing SVD-based coded technique. BICMB-FP is the special case of Bit-Interleaved Coded Multiple Beamforming with Constellation Precoding (BICMB-CP) whose system model is presented in Fig. \ref{fig:system_model_bicmb-cp}, when the number of precoded symbol streams equals to the number of employed subchannels.
To calculate one ML bit metric, ${1\over2}M^2$ constellation points are considered by exhaustive search, and the complexity is thereby $\mathcal{O}(M^{2})$. If SD presented in \cite{Azzam_SD_NLR}, \cite{Azzam_SD_RLR} is employed, the worst-case complexity for acquiring one ML bit metric is $\mathcal{O}(M)$ for both BICMB-GC and $2\times2$ BICMB-FP.

Moreover, similarly to the uncoded case, lower decoding complexity can be achieved for BICMB-GC because the $\mathbf{R}$ matrix in (\ref{eq:ML_bit_metrics_bicmb-pc_2})
is real-valued as proved in Section \ref{subsec:GCMB_Decoding}. As a result, the real and imaginary parts of $\tilde{\mathbf{y}}_{m,k}$ in (\ref{eq:ML_bit_metrics_bicmb-pc_2}) can be separated, and only the part corresponding to the coded bit is required for calculating one bit metric of the Viterbi decoder. Assume that square $M$-QAM is used, whose real and imaginary parts are Gray coded separately as two $\sqrt{M}$-PAM. Define $\Re [ \xi_{c_{k'}}^{n,j}]$ and $\Im [ \xi_{c_{k'}}^{n,j}]$ as the signal sets of the real and the imaginary axes of $\xi_{c_{k'}}^{n,j}$, respectively. Therefore, the ML bit metrics in (\ref{eq:ML_bit_metrics_bicmb-pc_2}) can be further simplified as
\begin{align}
\Gamma^{(m,n),j}(\mathbf{Y}_k, c_{k'}) = \min_{\Re[\mathbf{{x}}] \in \Re[\xi_{c_{k'}}^{n,j}]} \| \Re[\tilde{\mathbf{y}}_{m,k}] - \mathbf{R} \Re[\mathbf{x}] \|^2, \label{eq:ML_bit_metrics_real}
\end{align}
if the bit position of $c_{k'}$ is on the real part, or
\begin{align}
\Gamma^{(m,n),j}(\mathbf{Y}_k, c_{k'}) = \min_{\Im[\mathbf{{x}}] \in \Im[\xi_{c_{k'}}^{n,j}]} \| \Im[\tilde{\mathbf{y}}_{m,k}] - \mathbf{R} \Im[\mathbf{x}] \|^2, \label{eq:ML_bit_metrics_imag}
\end{align}
if the bit position of $c_{k'}$ is on the imaginary part.
For (\ref{eq:ML_bit_metrics_real}) and (\ref{eq:ML_bit_metrics_imag}), the worst-case complexity of acquiring one bit metric is only $\mathcal{O}(\sqrt{M})$ by using a real-valued SD with the last layer rounded, which is much lower than $\mathcal{O}(M)$ of $2\times2$ BICMB-FP.

\subsection{BICMB-PC} \label{subsec:BICMB-PC}
For BICMB-PC of dimension $D \in \{3,4,6\}$, it can be proved that they all achieve the full diversity order of $D^2$, which is generalized from 
(\ref{eq:PEP_average_bicmb-gc_2})-(\ref{eq:PEP_PSB_final_bicmb-gc}) because they are still valid for larger $D$.


For the decoding of BICMB-PC in dimension $D \in \{3,4,6\}$, similarly to BICMB-GC, the elements of $\mathbf{\Lambda} \mathbf{Z}_k$ are related to only one of the $\mathbf{x}_{v,k}$, thereby can be divided into $D$ groups, where the $v$th group contains elements related to $\mathbf{x}_{v,k}$. Then the received signal is divided into $D$ parts, which can be represented similarly to (\ref{eq:deteced_symbol_decomposed_pcmb}) as
\begin{align}
\mathbf{\breve{y}}_{v,k} = \mathbf{\Phi}_v \mathbf{\Lambda G} \mathbf{x}_{v,k} + \mathbf{\breve{n}}_{v,k}.
\label{eq:deteced_symbol_decomposed_bicmb-pc}
\end{align}
By using the QR decomposition of $\mathbf{\Lambda G}=\mathbf{Q} \mathbf{R}$, and moving $\mathbf{\Phi}_v\mathbf{Q}$ to the left hand, (\ref{eq:deteced_symbol_decomposed_bicmb-pc}) is rewritten as
\begin{align}
\tilde{\mathbf{y}}_{v,k} = \mathbf{Q}^H \mathbf{\Phi}_v^H \mathbf{\breve{y}}_{v,k} = \mathbf{R}\mathbf{x}_{v,k} + \mathbf{Q}^H \mathbf{\Phi}_v^H \mathbf{\breve{n}}_{v,k} = \mathbf{R}\mathbf{x}_{v,k} + \tilde{\mathbf{n}}_{v,k}.
\label{eq:deteced_symbol_decomposed_bicmb-pc_2}
\end{align}
Then the ML bit metrics in (\ref{eq:ML_bit_metrics_bicmb-pc}) can be simplified as (\ref{eq:ML_bit_metrics_bicmb-pc_2}).

In the case of $D=4$, the $\mathbf{R}$ matrix in (\ref{eq:ML_bit_metrics_bicmb-pc_2}) is real-valued. See the Appendix for the proof. As a result, the real and imaginary parts of $\mathbf{\tilde{y}}_{m,k}$ in (\ref{eq:ML_bit_metrics_bicmb-pc_2}) can be separated, and only the part corresponding to the coded bit is required for calculating one bit metric of the Viterbi decoder. Assume that square $M$-QAM is employed. Then, the ML bit metrics in (\ref{eq:ML_bit_metrics_bicmb-pc_2}) can be further simplified as (\ref{eq:ML_bit_metrics_real}) if the bit location of $c_{k'}$ is on the real part, or (\ref{eq:ML_bit_metrics_imag}) if the bit location of $c_{k'}$ is on the imaginary part. Therefore, the worst-case complexity for calculating one bit metric is only $\mathcal{O}(M^{1.5})$ by using a real-valued SD with the last layer rounded. On the other hand, BICMB-FP has the worst-case complexity of $\mathcal{O}(M^{3})$ by using a real-valued SD based on the real lattice representation in \cite{Azzam_SD_NLR},
\cite{Azzam_SD_RLR}, plus quantization of the last two layers.

For the dimension of $3$ ($6$), the $\mathbf{R}$ matrix is complex-valued. Therefore, the real and the imaginary parts of $\mathbf{\tilde{y}}_{m,k}$ cannot be separated, unlike the case of $D=2,4$. Moreover, since the M-HEX modulation is used instead of M-QAM, a complex-valued SD with an efficient implementation of a slicer \cite{Sinnokrot_STBC_LMLDC} is needed. The worst-case complexity of BICMB-PC to derive one bit metric is then $\mathcal{O}(M^{2})$ ($\mathcal{O}(M^{5})$). For BICMB-FP, the worst-case complexity is $\mathcal{O}(M^{2})$ ($\mathcal{O}(M^{5})$), which is similar to BICMB-PC.


\subsection{Discussion} \label{subsec:BICMB-PC_Discussions}

The worst-case complexity of BICMB-PC and BICMB-FP in different dimensions to calculate one bit metric is also summarized in Table \ref{table:complexity}. Note that they are actually the same as PCMB and FPMB.

Similarly, CSIT is assumed to be known for both BICMB-PC and BICMB-FP, and both of them achieve full diversity and full multiplexing. In dimensions $3$ and $6$, the worst-case complexity of BICMB-PC is similar to BICMB-FP. However, the real-valued $\mathbf{R}$ matrices in (\ref{eq:deteced_symbol_decomposed_bicmb-gc_2}) and (\ref{eq:deteced_symbol_decomposed_bicmb-pc_2}) cause the complexity advantages of BICMB-PC in dimensions $2$ and $4$ over BICMB-FP, whose $\mathbf{R}$ matrices are complex-valued. In this case, the $D$-dimensional complex-valued metric calculation problem can be decomposed into only one $D$-dimensional real-valued problem, instead of two $D$-dimensional real-valued problems for the uncoded case, because only one of the real and imaginary parts which corresponds to the coded bit needs to be considered. Therefore, the real-valued $\mathbf{R}$ matrices benefit BICMB-PC more than PCMB.

For the constellation precoding technique \cite{Park_CPB}, \cite{Park_CPMB}, unlike the uncoded case where only full precoding of FPMB can achieve both full diversity and full multiplexing, partial precoding of BICMB-PP for the coded case could also achieve both of them \cite{Park_CPMB}. Note that the precoded part of BICMB-PP could be considered as a smaller dimensional BICMB-FP. Therefore, BICMB-PC of dimensions $2$ and $4$ could be applied to replace the precoded part of BICMB-PP and reduce its decoding complexity. 

%% file: Results.tex
\section{Results} \label{sec:Results}
\subsection{PCMB} \label{subsec:Results_PCMB}
As presented Section \ref{sec:PCMB}, PCMB in dimensions $2$ and $4$ has the most advantage in terms of decoding complexity. Therefore. simulations are focused on these two dimensions. 

Considering $2\times2$ systems, Fig. \ref{fig:ber_snr_2x2_pcmb} shows BER-SNR performance comparison of GCMB, FPMB, and a MIMO system using GC, which is denoted by GC, for different modulation schemes. The constellation precoder for FPMB is selected as the best one introduced in \cite{Park_CPB}. Simulation results show that GC, FPMB, and GCMB, with the worst-case decoding complexity of $\mathcal{O}(M^{2.5})$, $\mathcal{O}(M)$, and $\mathcal{O}(\sqrt{M})$, respectively,
achieve very close performance for all of $4$-QAM, $16$-QAM, and $64$-QAM. The performance differences among these three are less than $1$dB, and become smaller when the modulation alphabet size increases. In fact, the performance loss mentioned in Section \ref{subsec:PCMB_Discussions} is negligible in the $2\times2$ case.

In the case of $4\times4$ systems, Fig. \ref{fig:ber_snr_4x4_pcmb} shows BER-SNR performance comparison of PCMB, FPMB, PC, for $4$-QAM and $16$-QAM. The constellation precoder for FPMB is also chosen as the best one in \cite{Park_CPB}. Simulation results show that PCMB has approximately $3$dB and $1$dB performance degradations compared to PC and FPMB, respectively, and the degradations decrease as the modulation alphabet size increases. However, the modest performance compromises of PCMB in the $4\times4$ case trade off with substantial reductions of the worst-case decoding complexity for PC and FPMB from $\mathcal{O}(M^{13.5})$ and $\mathcal{O}(M^{3})$ to only $\mathcal{O}(M^{1.5})$, respectively. 

Obviously, the execution of SD with lower dimension has less complexity, including the worst case and the average case. Therefore,  the worst-case decoding complexity is used to roughly compare the complexity of PC, FPMB, and PCMB in this paper above. In order to measure the average decoding complexity and show the exact complexity comparisons, the average number of real multiplications, which are the most expensive operations in terms of machine cycles, for decoding one transmitted vector symbol are calculated at different SNR for PC, FPMB, and PCMB, respectively. In \cite{Li_RCSD_J}, \cite{Li_RCSD}, a reduced complexity SD technique substantially decreasing the average number of real multiplications was introduced, which is employed in this paper. Fig. \ref{fig:complexity_gcmb1} and Fig. \ref{fig:complexity_gcmb2} show the complexity comparisons of GCMB with GC and FPMB respectively, for $2\times2$ MIMO systems using $64$-QAM. The complexity of GCMB is $99\%$ and $48\%$ 
lower than GC at low and high SNR respectively, while it is $70\%$ 
lower than FPMB at low SNR and close to FPMB at high SNR. Fig. \ref{fig:complexity_pcmb1} shows the complexity comparisons for PCMB and PC for $4\times4$ MIMO systems using $4$-QAM. The complexity of PCMB is $2.7$ and $1.7$ orders of magnitude lower than PC at low and high SNR respectively. Fig. \ref{fig:complexity_pcmb2} shows the complexity comparisons for PCMB and FPMB for $4\times4$ MIMO systems using $16$-QAM. The complexity of PCMB is $85\%$ 
lower than FPMB at low SNR and close to FPMB at high SNR. Note that the improvements will be much greater for larger alphabet size.

\ifCLASSOPTIONonecolumn
\begin{figure}[!m]
\centering
\centering \includegraphics[width = 0.6\linewidth]{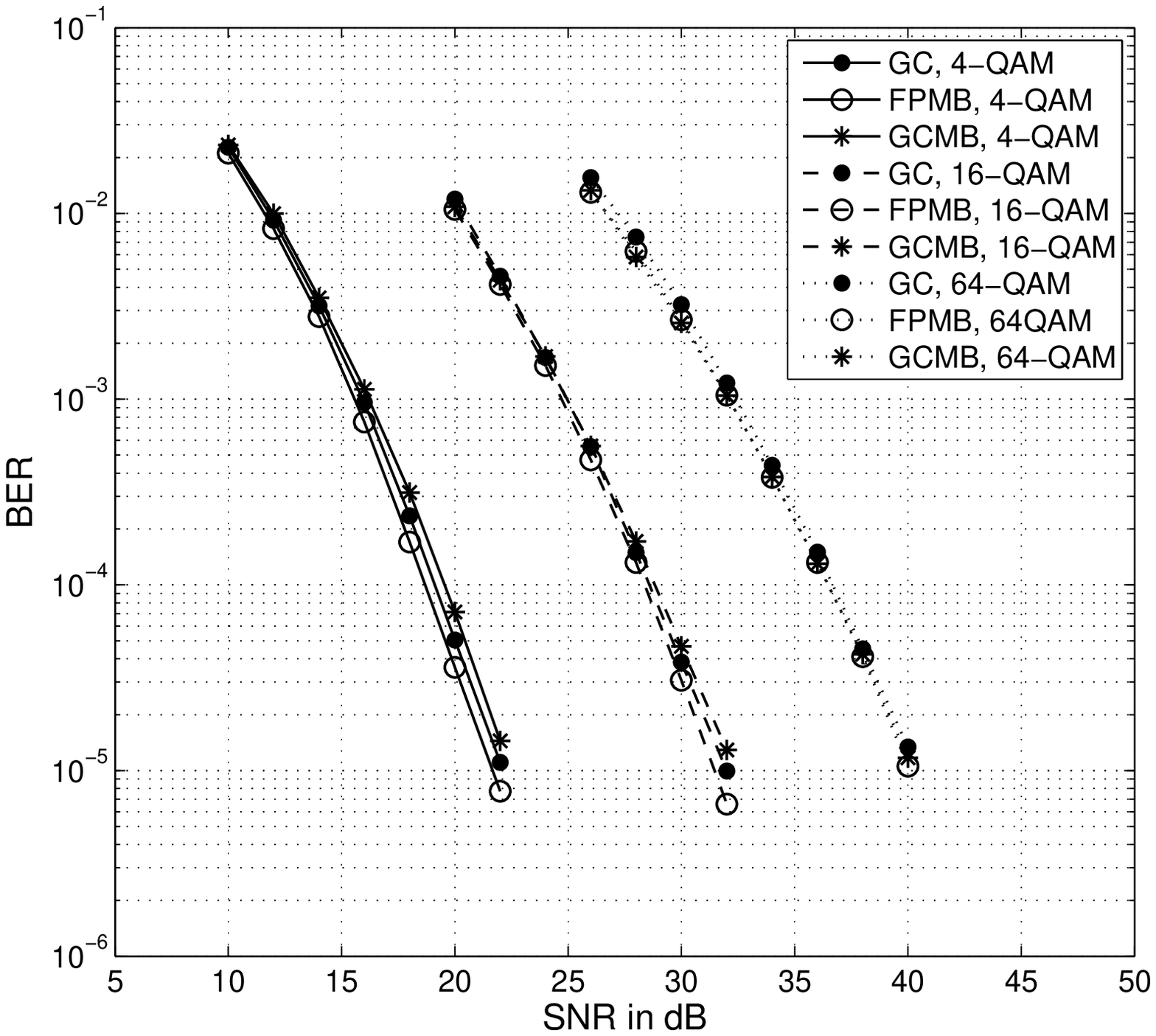}
\caption{BER vs. SNR of GC, FPMB, and GCMB for $2\times2$ systems.}
\label{fig:ber_snr_2x2_pcmb}
\end{figure}

\begin{figure}[!m]
\centering \includegraphics[width = 0.6\linewidth]{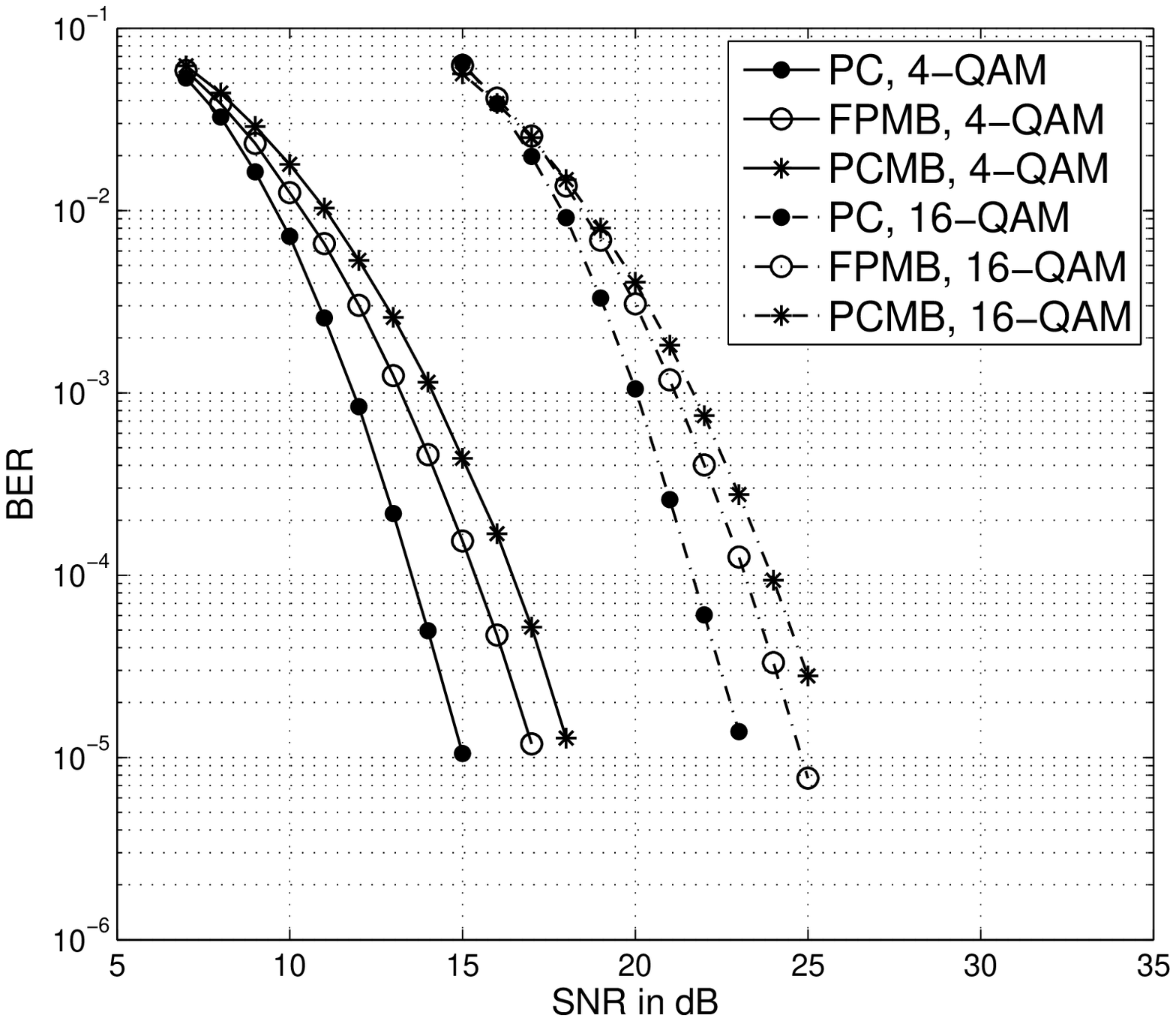}
\caption{BER vs. SNR of PC, FPMB, and PCMB for $4\times4$ systems.}
\label{fig:ber_snr_4x4_pcmb}
\end{figure}

\begin{figure}[!m]
\centering \includegraphics[width = 0.6\linewidth]{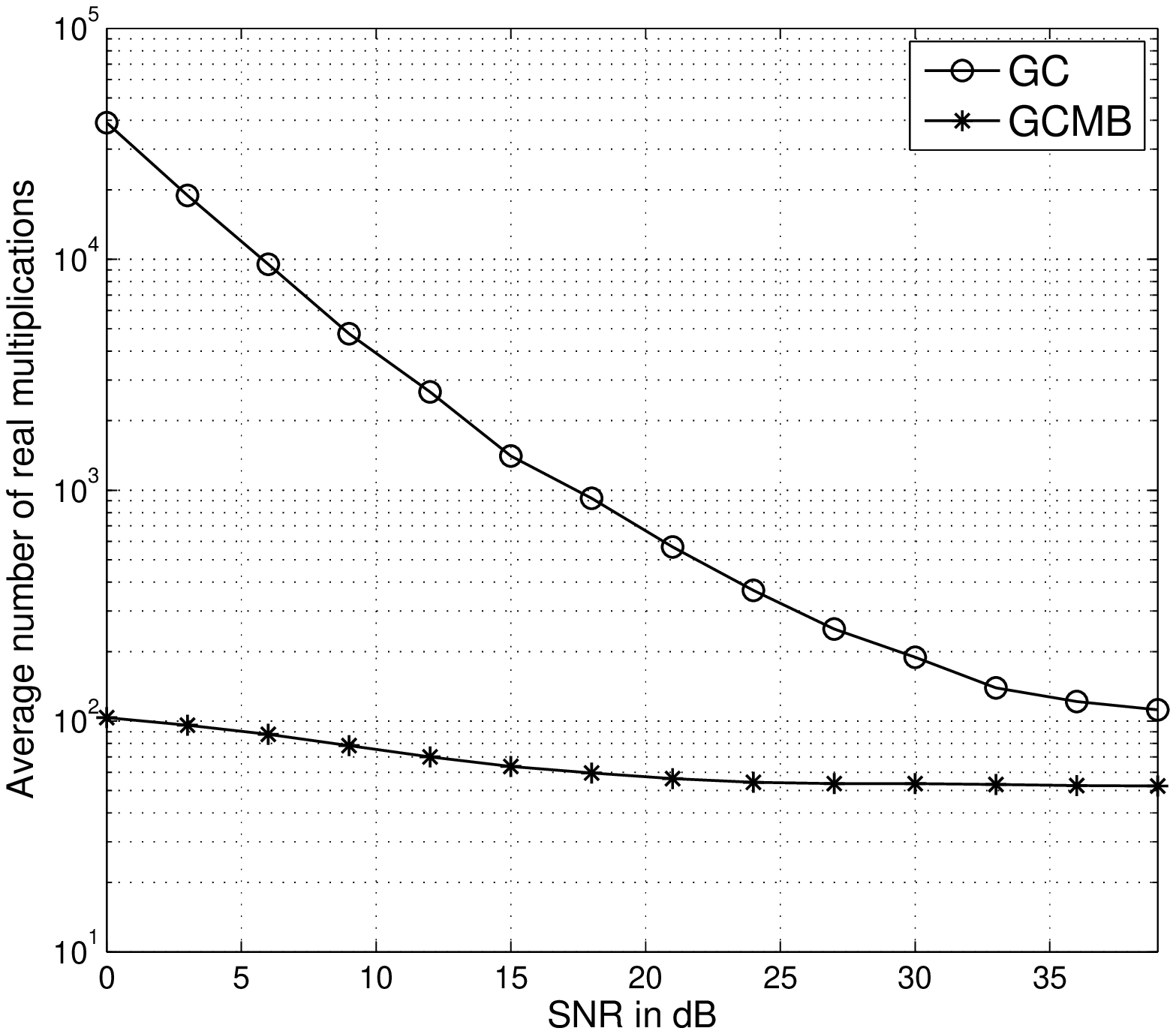}
\caption{Average number of real multiplications vs. SNR of GC and GCMB for $2\times2$ systems using $64$-QAM.}
\label{fig:complexity_gcmb1}
\end{figure}

\begin{figure}[!m]
\centering \includegraphics[width = 0.6\linewidth]{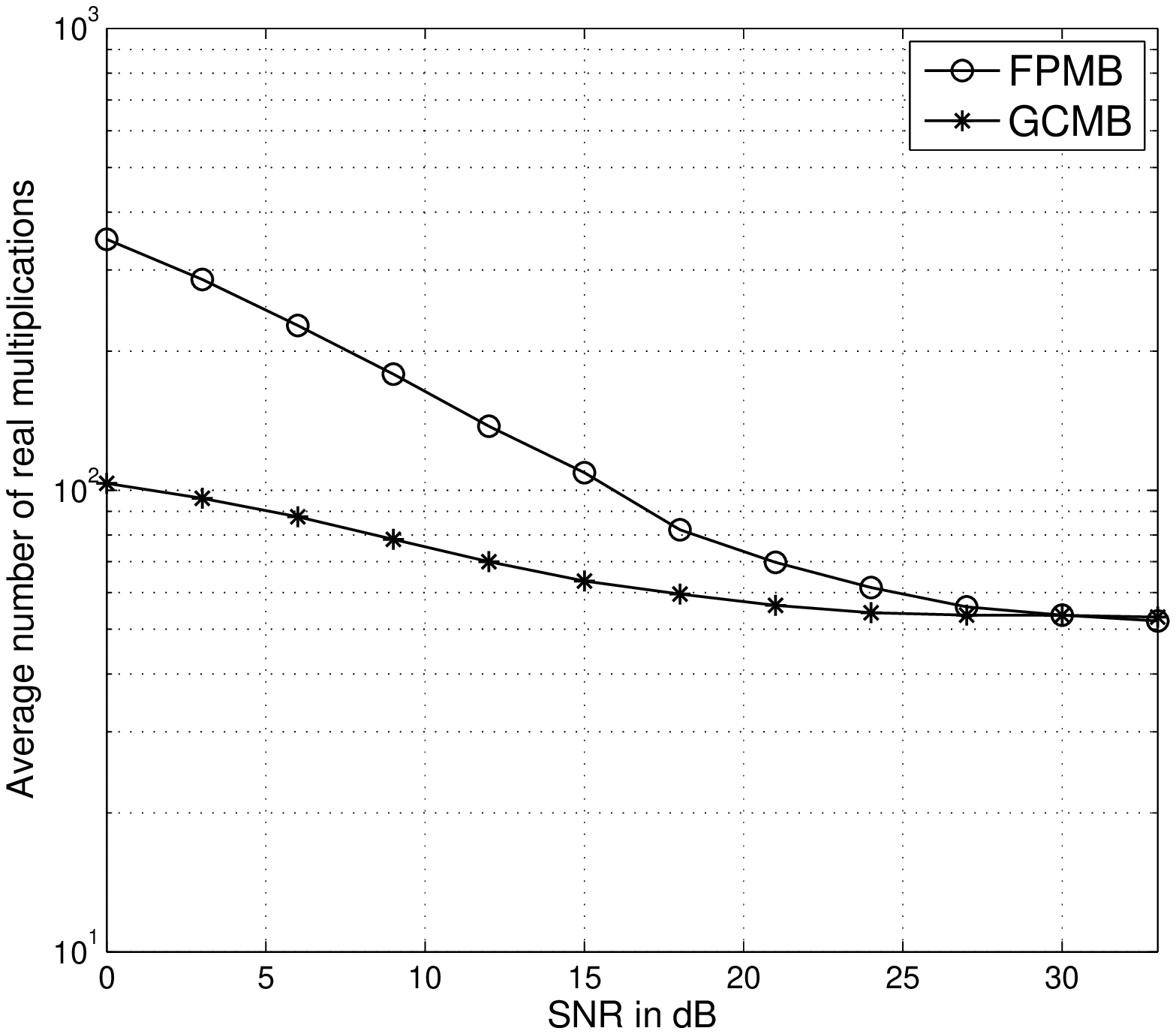}
\caption{Average number of real multiplications vs. SNR of FPMB and GCMB for $2\times2$ systems using $64$-QAM.}
\label{fig:complexity_gcmb2}
\end{figure}

\begin{figure}[!m]
\centering \includegraphics[width = 0.6\linewidth]{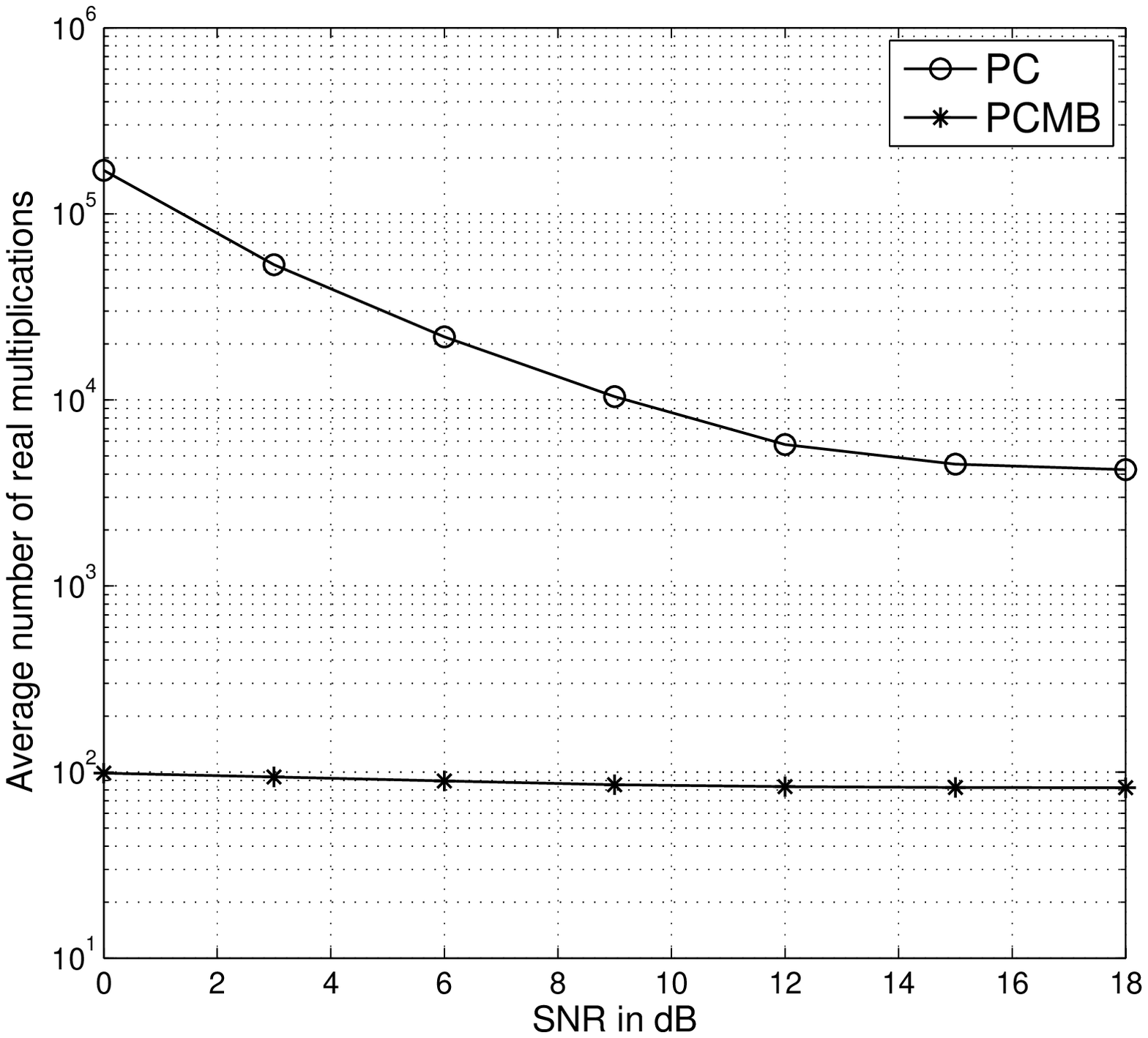}
\caption{Average number of real multiplications vs. SNR of PC and PCMB for $4\times4$ systems using $4$-QAM.}
\label{fig:complexity_pcmb1}
\end{figure}

\begin{figure}[!m]
\centering \includegraphics[width = 0.6\linewidth]{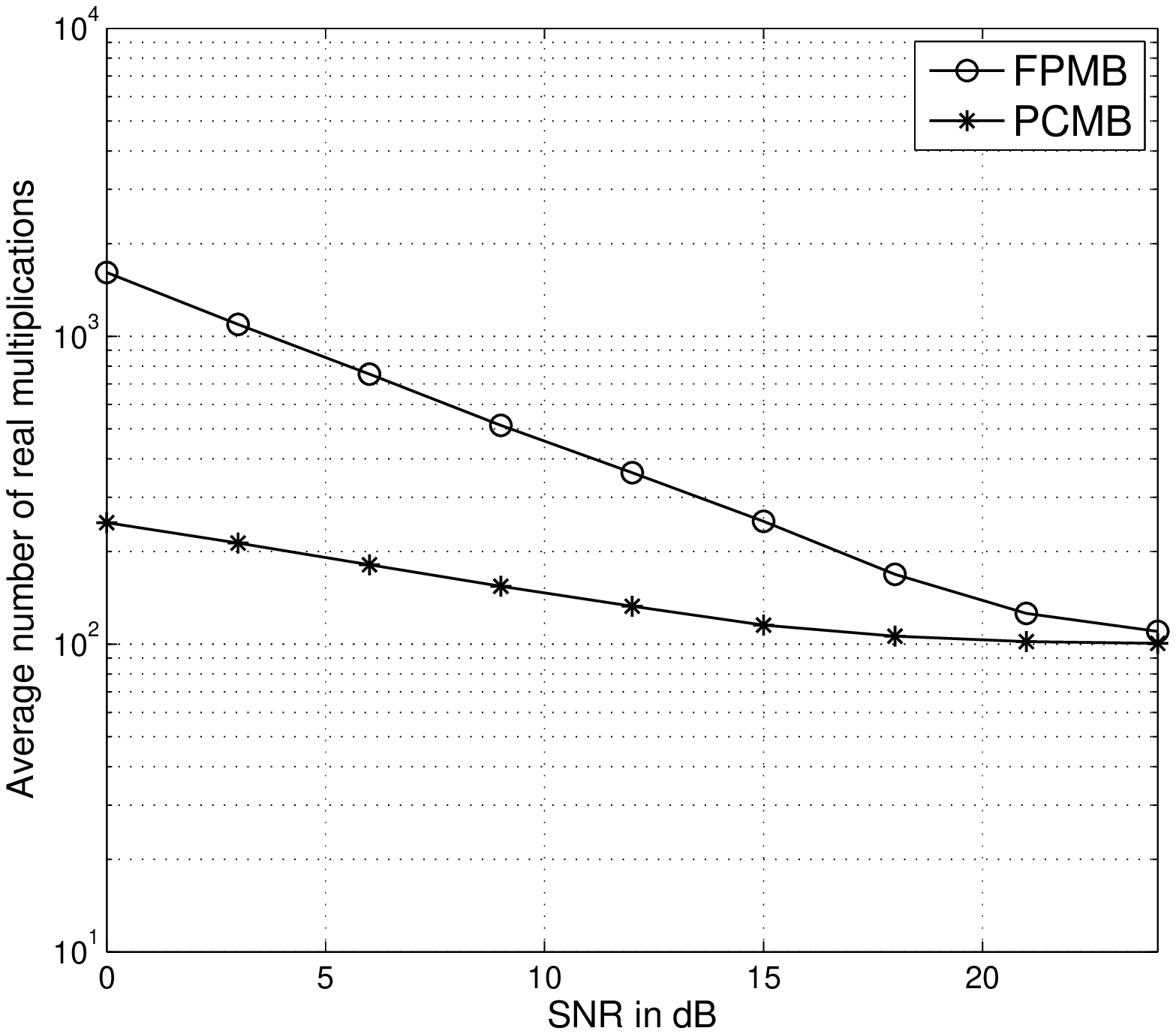}
\caption{Average number of real multiplications vs. SNR for FPMB and PCMB for $4\times4$ systems using $16$-QAM.}
\label{fig:complexity_pcmb2}
\end{figure}

\else
\begin{figure}[!t]
\centering \includegraphics[width = \sizefig\linewidth]{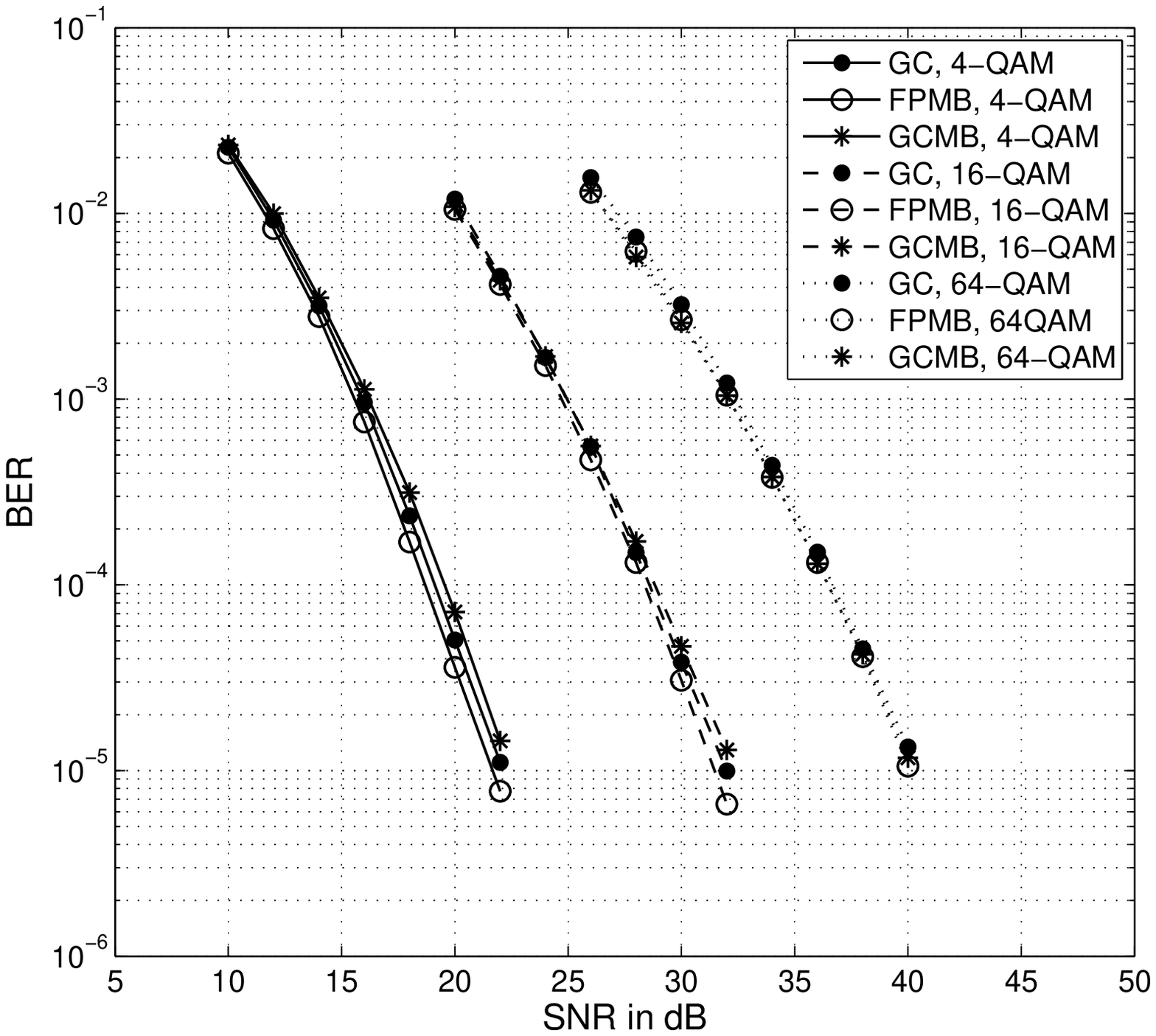}
\caption{BER vs. SNR of GC, FPMB, and GCMB for $2\times2$ systems.}
\label{fig:ber_snr_2x2_pcmb}
\end{figure}

\begin{figure}[!t]
\centering \includegraphics[width = \sizefig\linewidth]{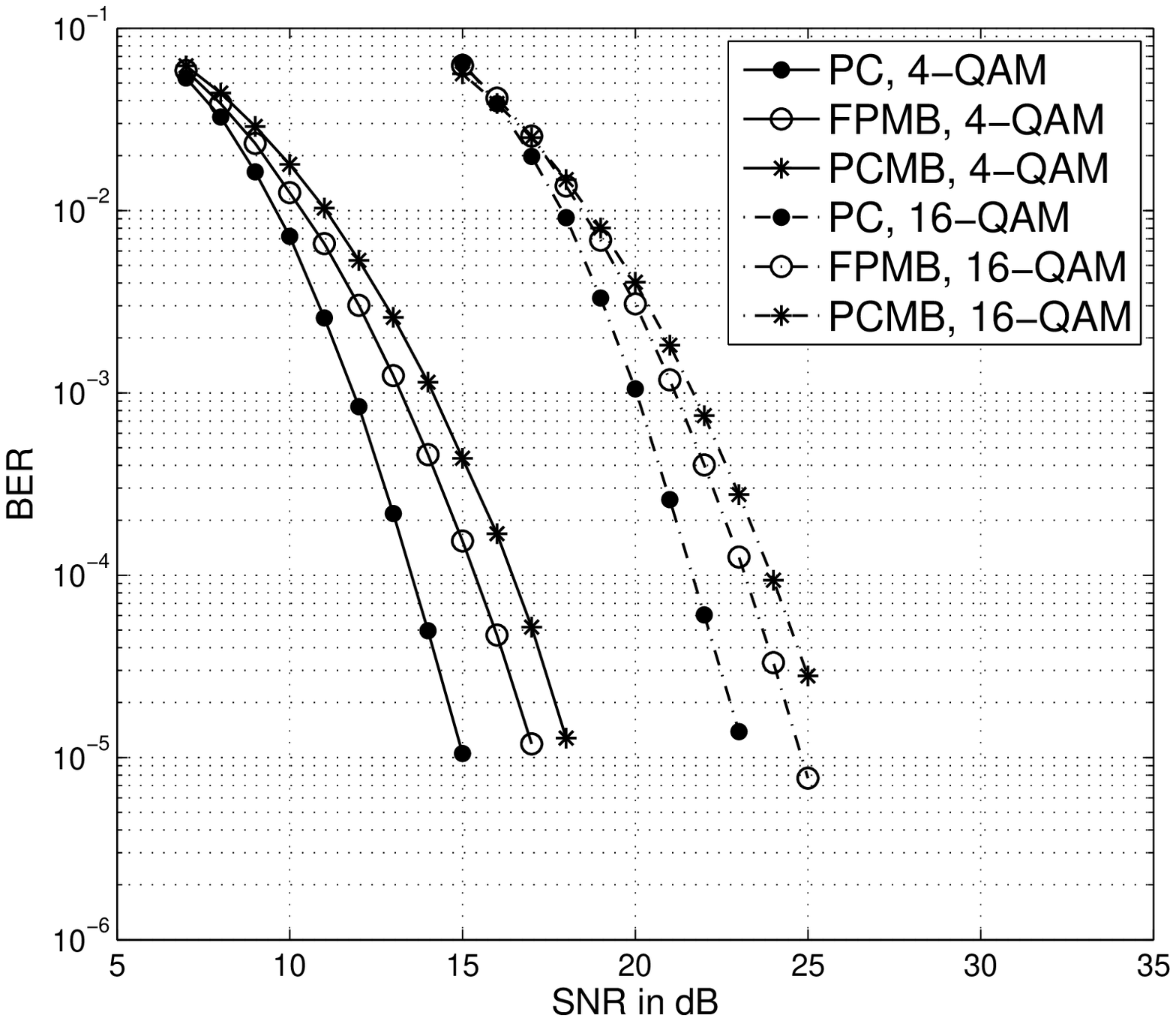}
\caption{BER vs. SNR of PC, FPMB, and PCMB for $4\times4$ systems.}
\label{fig:ber_snr_4x4_pcmb}
\end{figure}

\begin{figure}[!t]
\centering \includegraphics[width = \sizefig\linewidth]{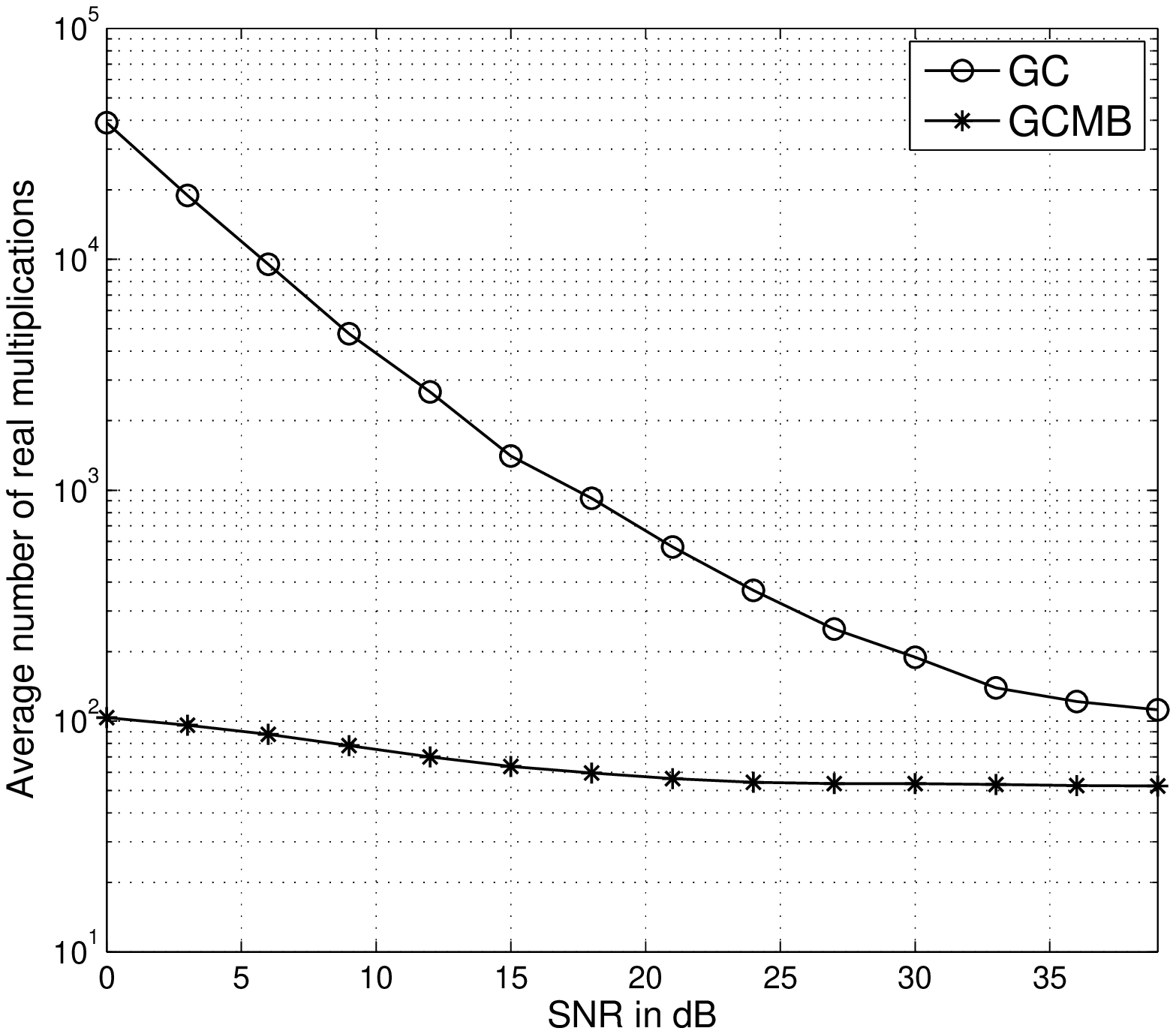}
\caption{Average number of real multiplications vs. SNR of GC and GCMB for $2\times2$ systems using $64$-QAM.}
\label{fig:complexity_gcmb1}
\end{figure}

\begin{figure}[!t]
\centering \includegraphics[width = \sizefig\linewidth]{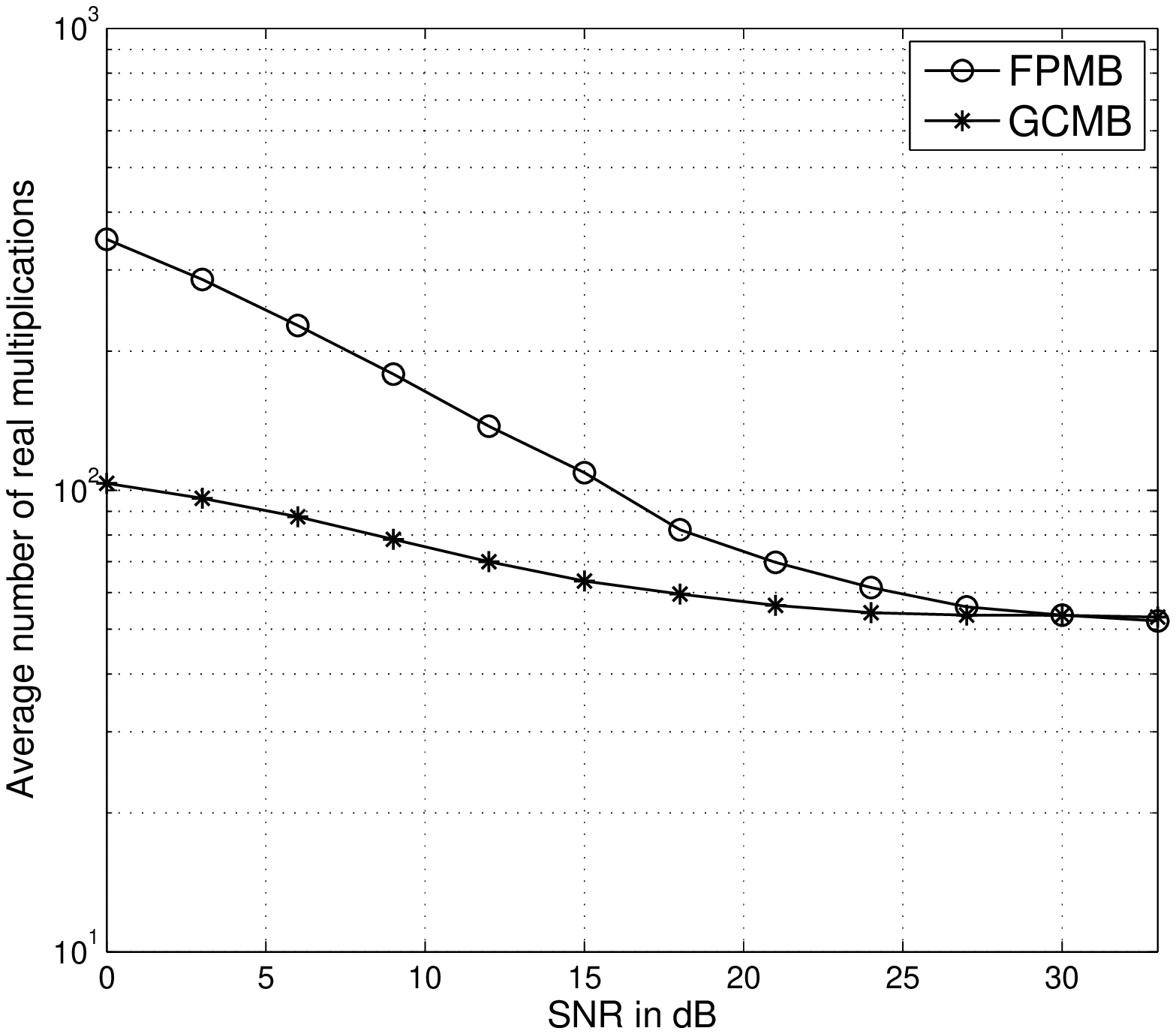}
\caption{Average number of real multiplications vs. SNR of FPMB and GCMB for $2\times2$ systems using $64$-QAM.}
\label{fig:complexity_gcmb2}
\end{figure}

\begin{figure}[!t]
\centering \includegraphics[width = \sizefig\linewidth]{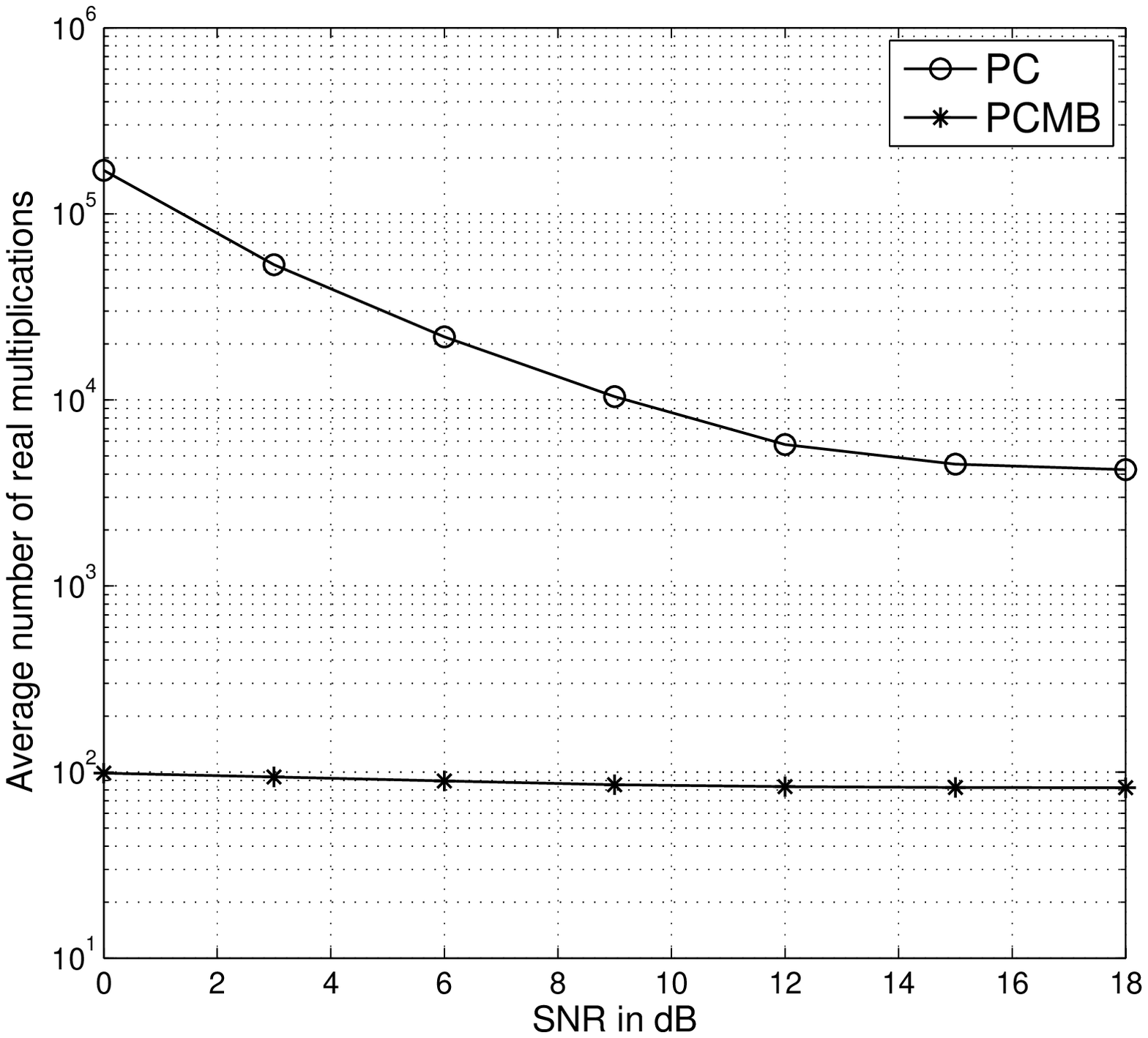}
\caption{Average number of real multiplications vs. SNR of PC and PCMB for $4\times4$ systems using $4$-QAM.}
\label{fig:complexity_pcmb1}
\end{figure}

\begin{figure}[!t]
\centering \includegraphics[width = \sizefig\linewidth]{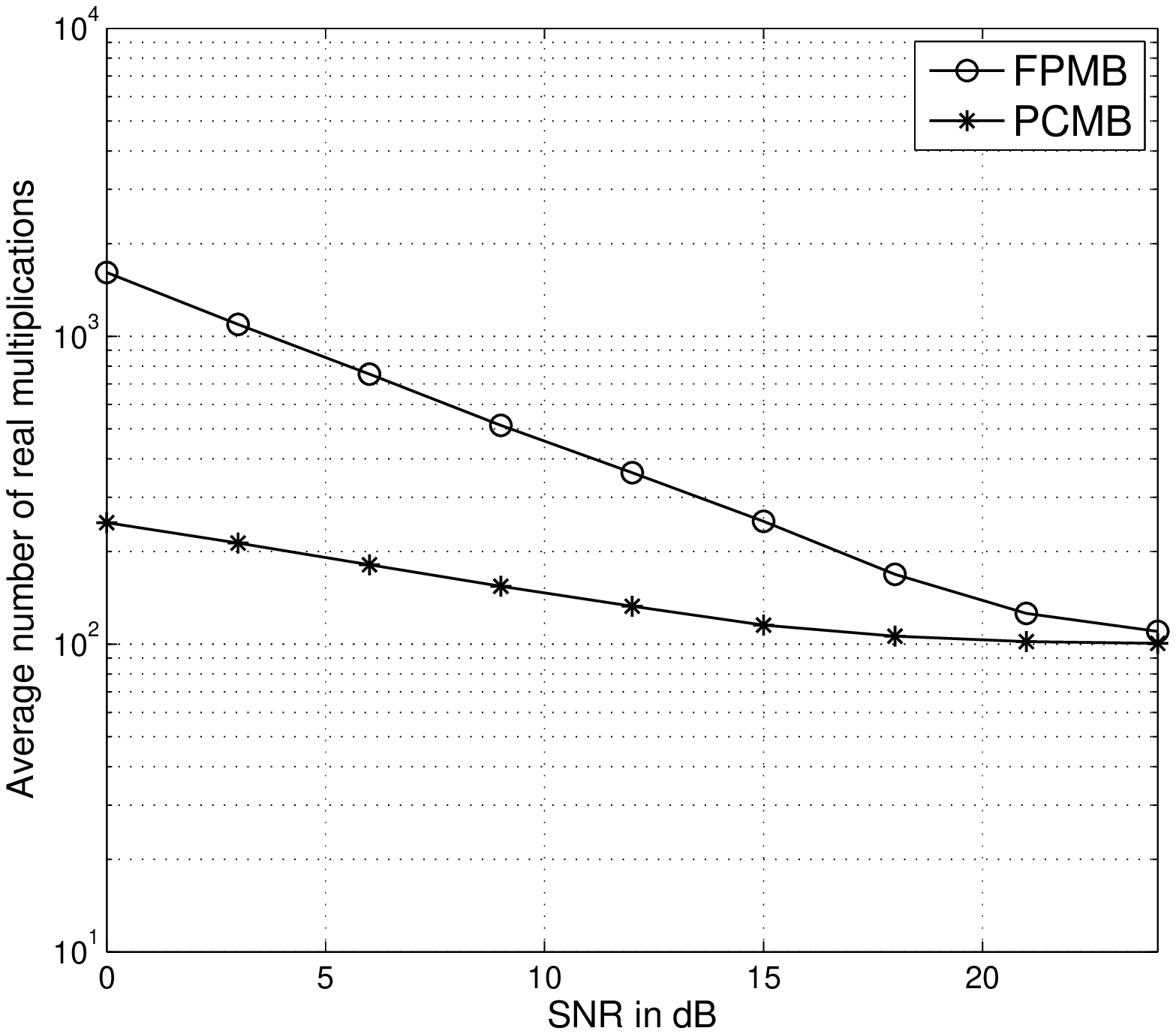}
\caption{Average number of real multiplications vs. SNR for FPMB and PCMB for $4\times4$ systems using $16$-QAM.}
\label{fig:complexity_pcmb2}
\end{figure}

\fi

\subsection{BICMB-PC}
As presented Section \ref{sec:BICMB-PC}, BICMB-PC in dimensions $2$ and $4$ has the most advantage in terms of complexity for calculating bit metrics. Therefore. simulations are focused on these two dimensions. 

Considering $R_c=2/3$, $2\times2$ systems, Fig. \ref{fig:ber_snr_2x2_bicmb-gc} shows BER-SNR performance comparison of BICMB-PC and BICMB-FP. The constellation precoder for BICMB-FP is selected as the best one introduced in \cite{Park_CPB}. Simulation results show that BICMB-FP and BICMB-PC, with the worst-case decoding complexity of $\mathcal{O}(M)$ and $\mathcal{O}(\sqrt{M})$ to acquire one bit metric respectively, achieve almost the same performance for all of $4$-QAM, $16$-QAM, and $64$-QAM. 

In the case of $R_c=4/5$, $4\times4$ systems, Fig. \ref{fig:ber_snr_4x4_bicmb-pc} shows BER-SNR performance comparison of BICMB-PC and BICMB-FP for $4$-QAM and $16$-QAM. The constellation precoder for BICMB-FP is also chosen as the best one in \cite{Park_CPB}. Similarly, simulation results show that BICMB-PC achieves almost the same performance as BICMB-FP. Moreover, the worst-case complexity of $\mathcal{O}(M^{1.5})$ to get one bit metric for BICMB-PC is much lower than that of $\mathcal{O}(M^3)$ for 
BICMB-FP. 

In order to measure the average decoding complexity, the average number of real multiplications for acquiring one bit metric is calculated at different SNR for BICMB-FP and BICMB-PC. In \cite{Li_RC_BICMB_CP}, an efficient reduced complexity decoding technique was introduced for BICMB-FP, which is applied in this paper. For fair comparisons, a similar decoding technique is employed to BICMB-PC. Fig. \ref{fig:complexity_bicmb-gc} shows the complexity comparisons for BICMB-PC and BICMB-FP for $2\times2$ MIMO systems using $64$-QAM. The complexity of BICMB-PC is $86\%$ and $70\%$ 
lower than BICMB-FP at low and high SNR respectively. Fig. \ref{fig:complexity_bicmb-pc} shows the complexity comparisons for BICMB-PC and BICMB-FP for $4\times4$ MIMO systems using $16$-QAM. The complexity of BICMB-PC is $1.7$ and $1.3$ orders of magnitude lower than BICMB-FP at low and high SNR respectively. Note that the number of improvements will be much greater for larger alphabet size.

\ifCLASSOPTIONonecolumn
\begin{figure}[!m]
\centering \includegraphics[width = 0.6\linewidth]{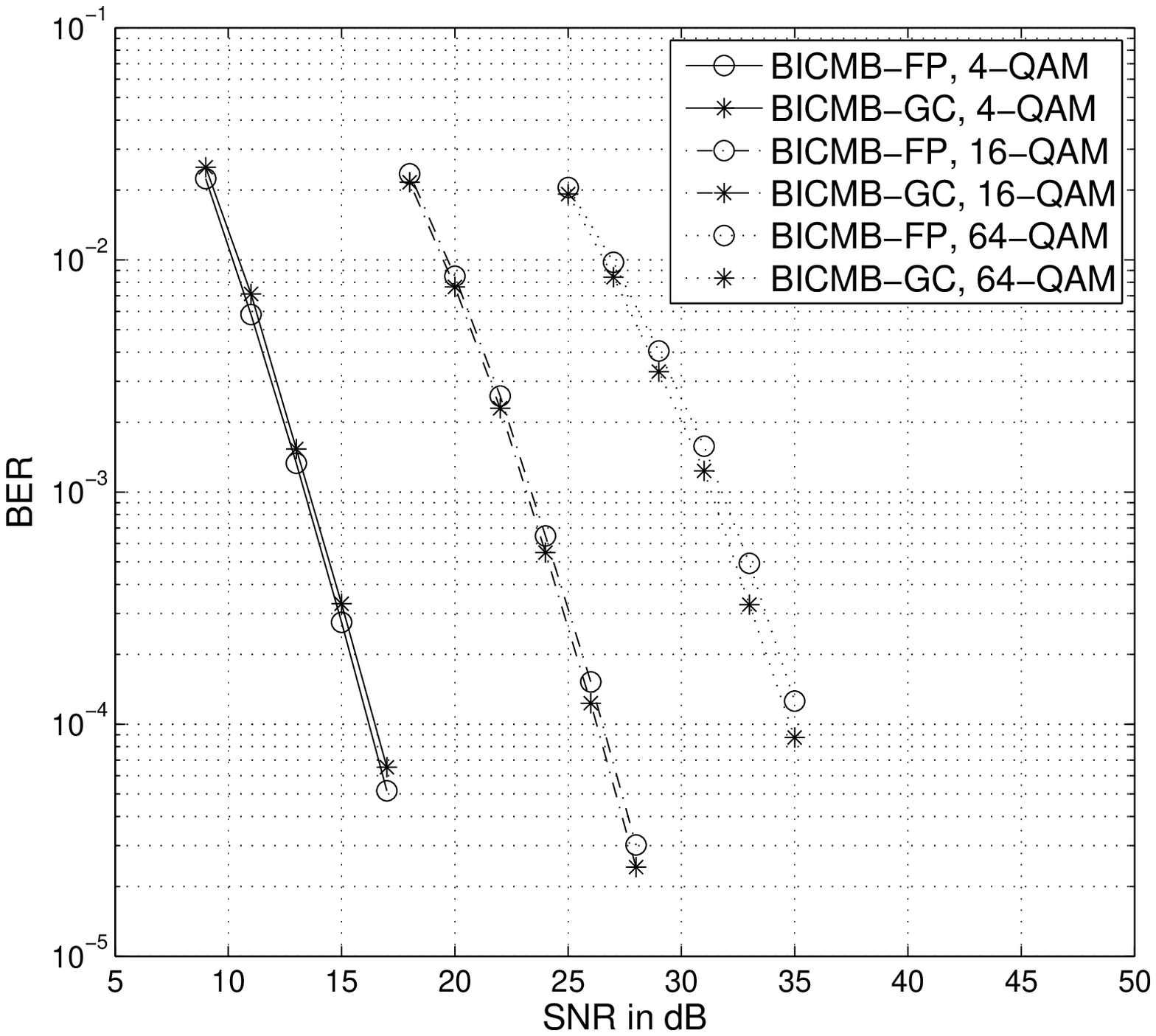}
\caption{BER vs. SNR of BICMB-FP and BICMB-GC for $R_c=2/3$, $2\times2$ systems.}
\label{fig:ber_snr_2x2_bicmb-gc}
\end{figure}

\begin{figure}[!m]
\centering \includegraphics[width = 0.6\linewidth]{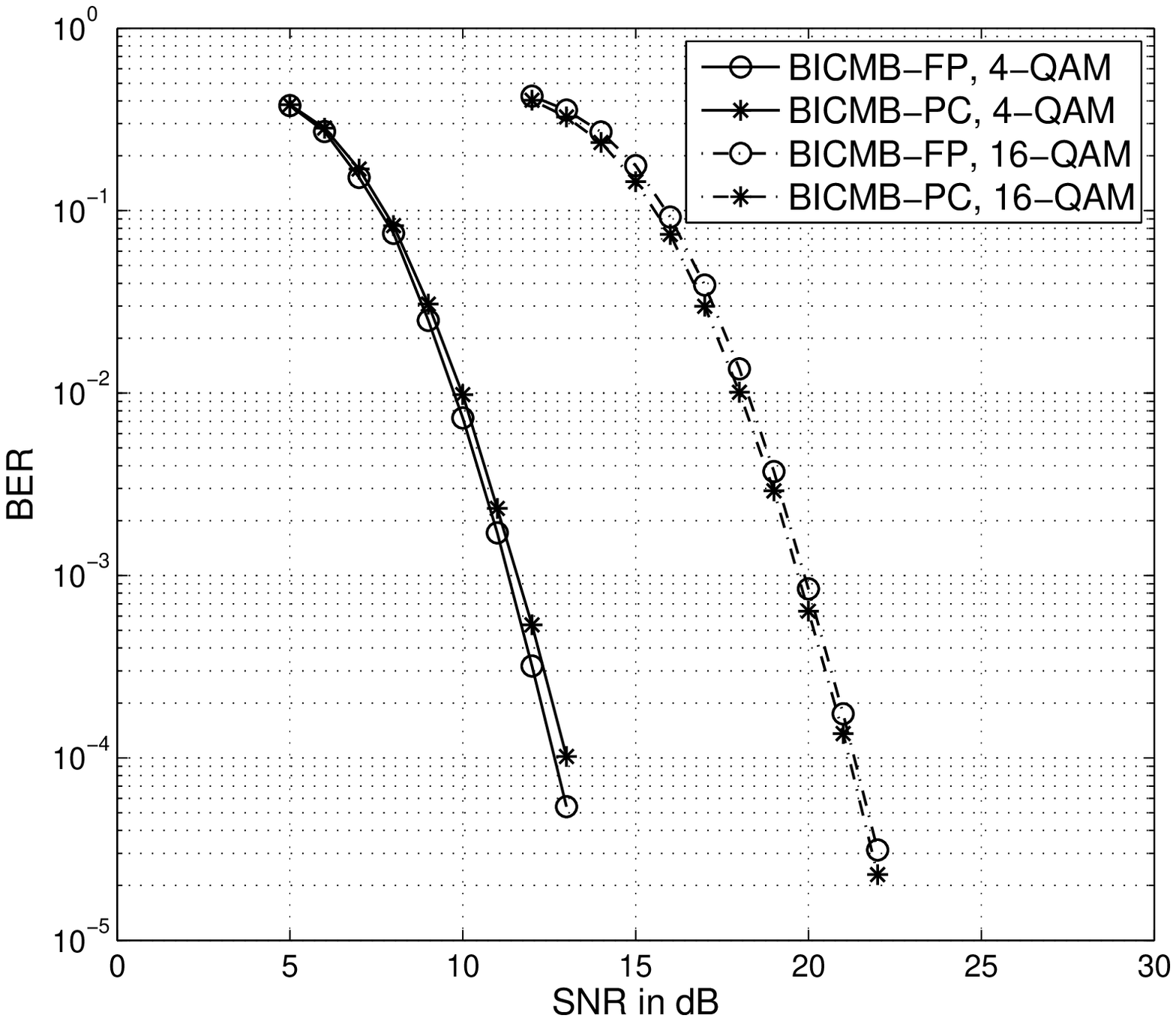}
\caption{BER vs. SNR of BICMB-FP and BICMB-PC for $R_c=4/5$, $4\times4$ systems.}
\label{fig:ber_snr_4x4_bicmb-pc}
\end{figure}

\begin{figure}[!m]
\centering \includegraphics[width = 0.6\linewidth]{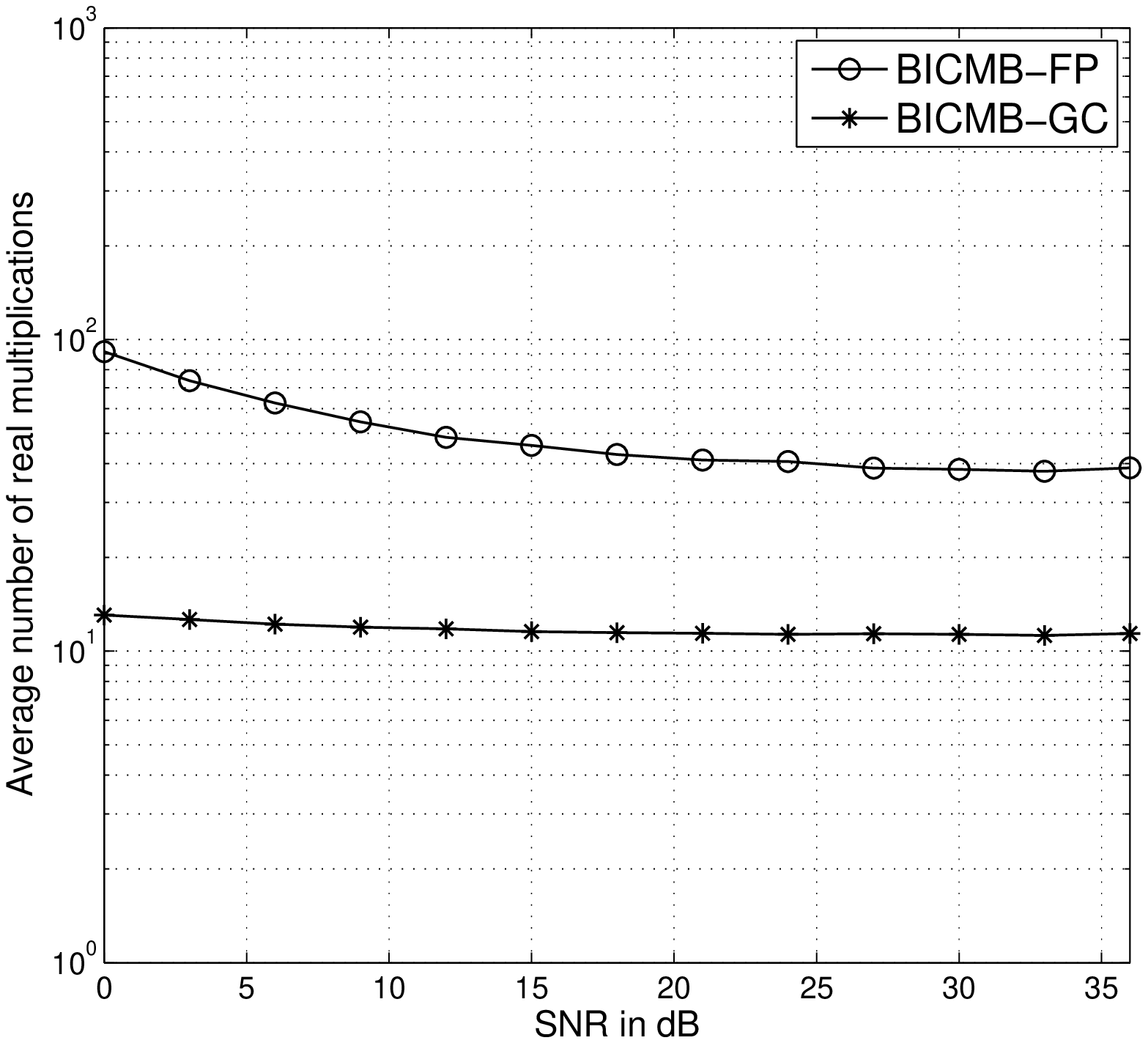}
\caption{Average number of real multiplications vs. SNR of BICMB-FP and BICMB-GC for $R_c=2/3$, $2\times2$ systems using $64$-QAM.}
\label{fig:complexity_bicmb-gc}
\end{figure}

\begin{figure}[!m]
\centering \includegraphics[width = 0.6\linewidth]{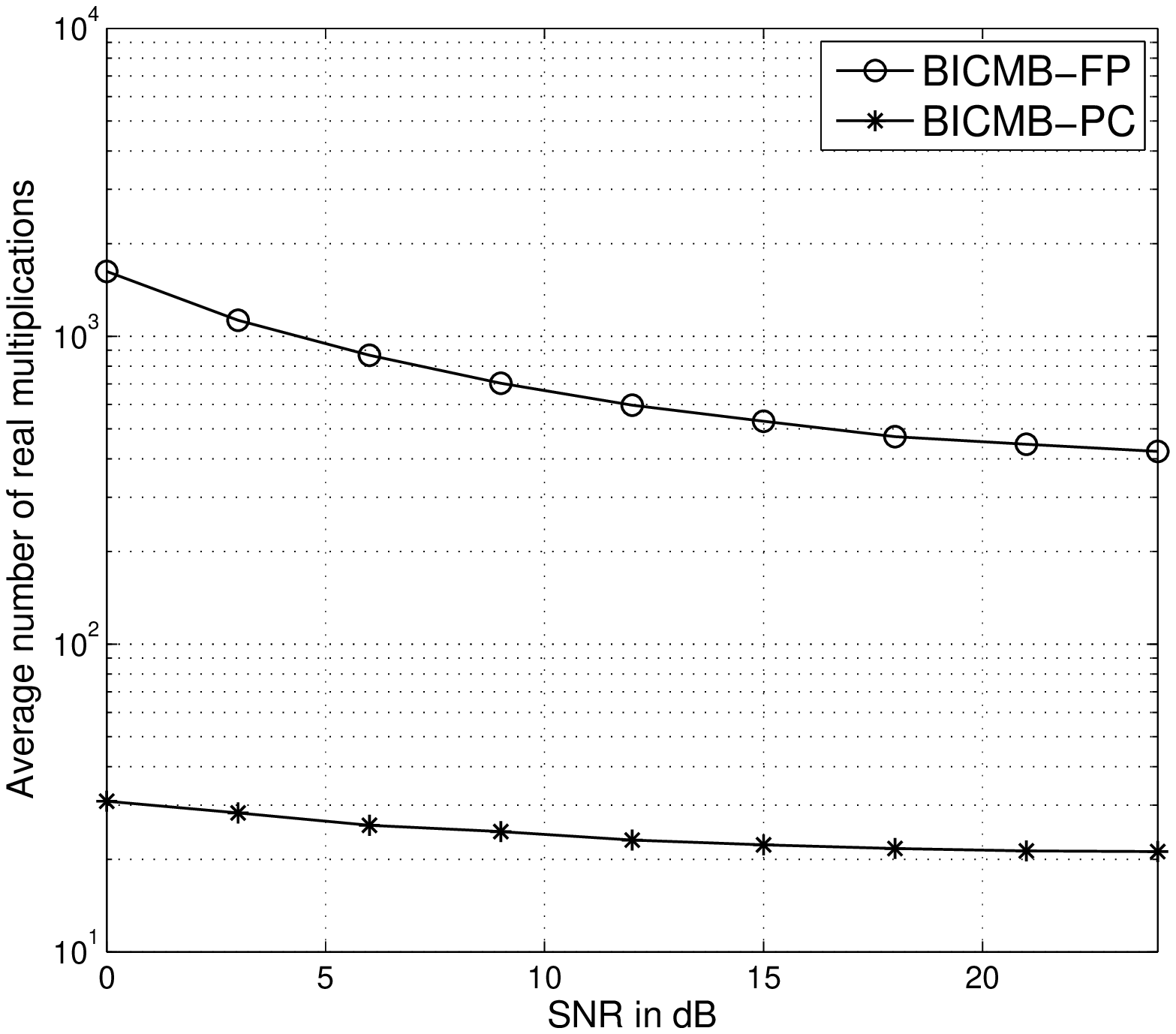}
\caption{Average number of real multiplications vs. SNR of BICMB-FP and BICMB-PC for $R_c=4/5$, $4\times4$ systems using $16$-QAM.}
\label{fig:complexity_bicmb-pc}
\end{figure}

\else
\begin{figure}[!t]
\centering \includegraphics[width = \sizefig\linewidth]{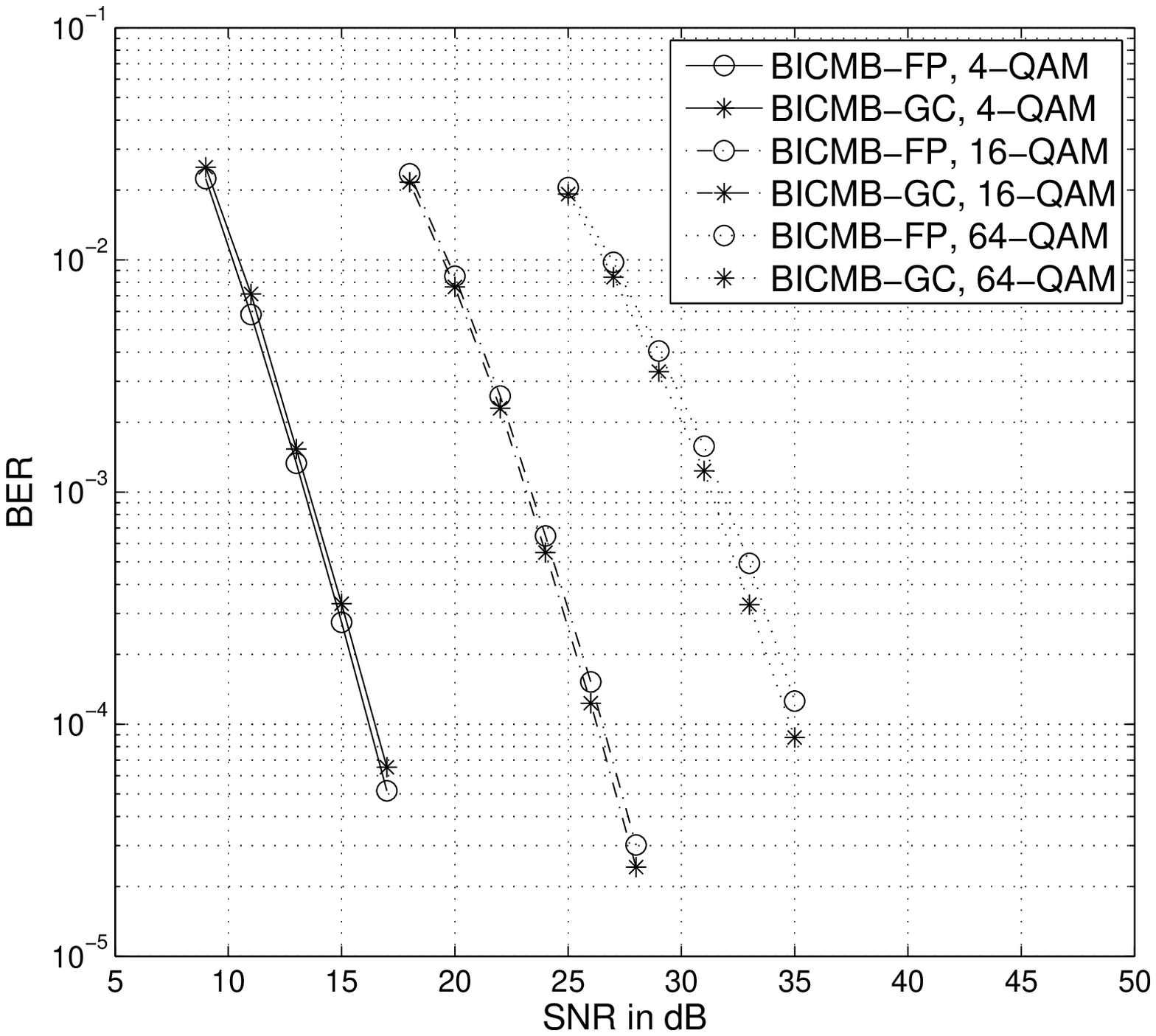}
\caption{BER vs. SNR of BICMB-FP and BICMB-GC for $R_c=2/3$, $2\times2$ systems.}
\label{fig:ber_snr_2x2_bicmb-gc}
\end{figure}

\begin{figure}[!t]
\centering \includegraphics[width = \sizefig\linewidth]{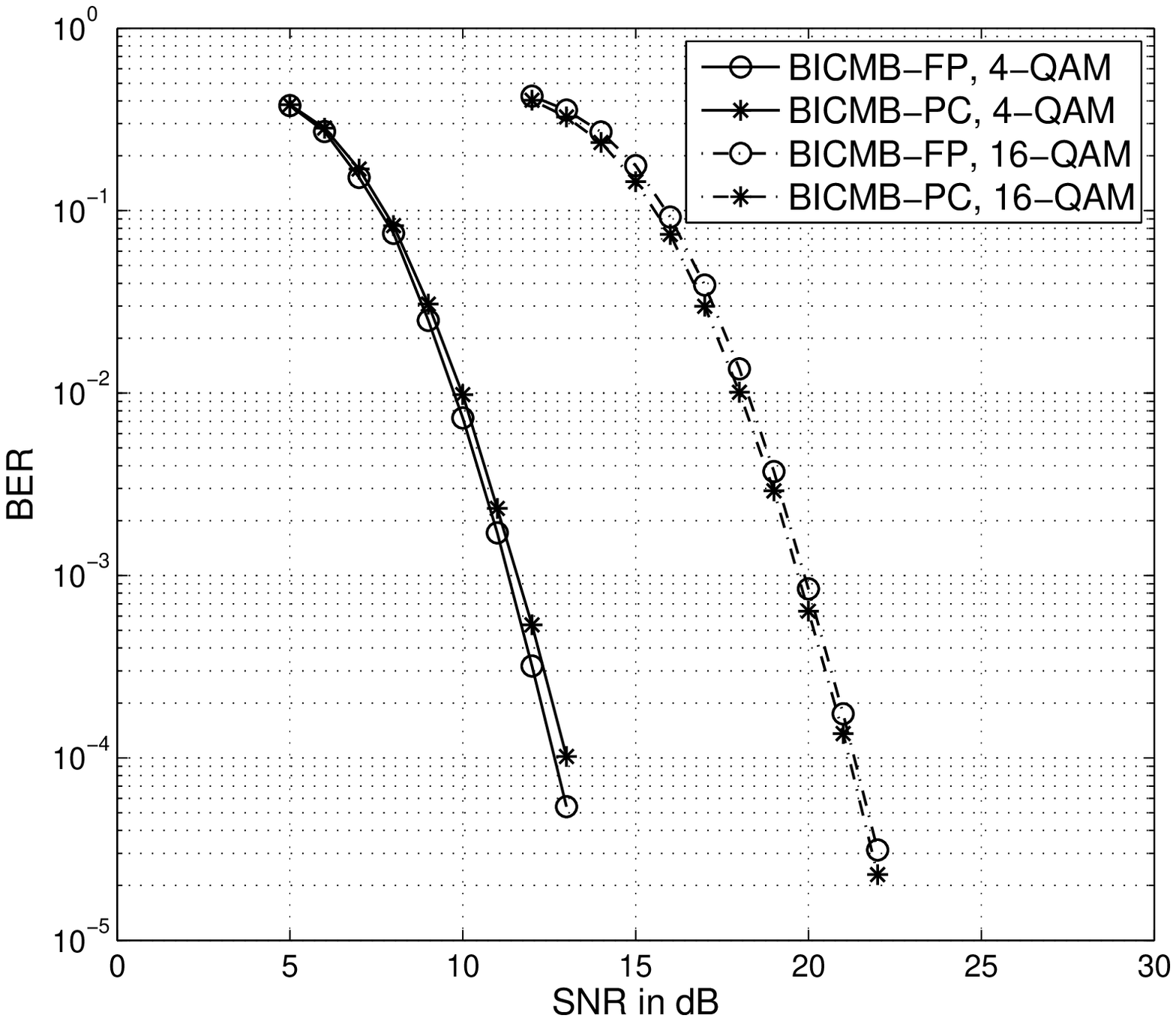}
\caption{BER vs. SNR of BICMB-FP and BICMB-PC for $R_c=4/5$, $4\times4$ systems.}
\label{fig:ber_snr_4x4_bicmb-pc}
\end{figure}

\begin{figure}[!t]
\centering \includegraphics[width = \sizefig\linewidth]{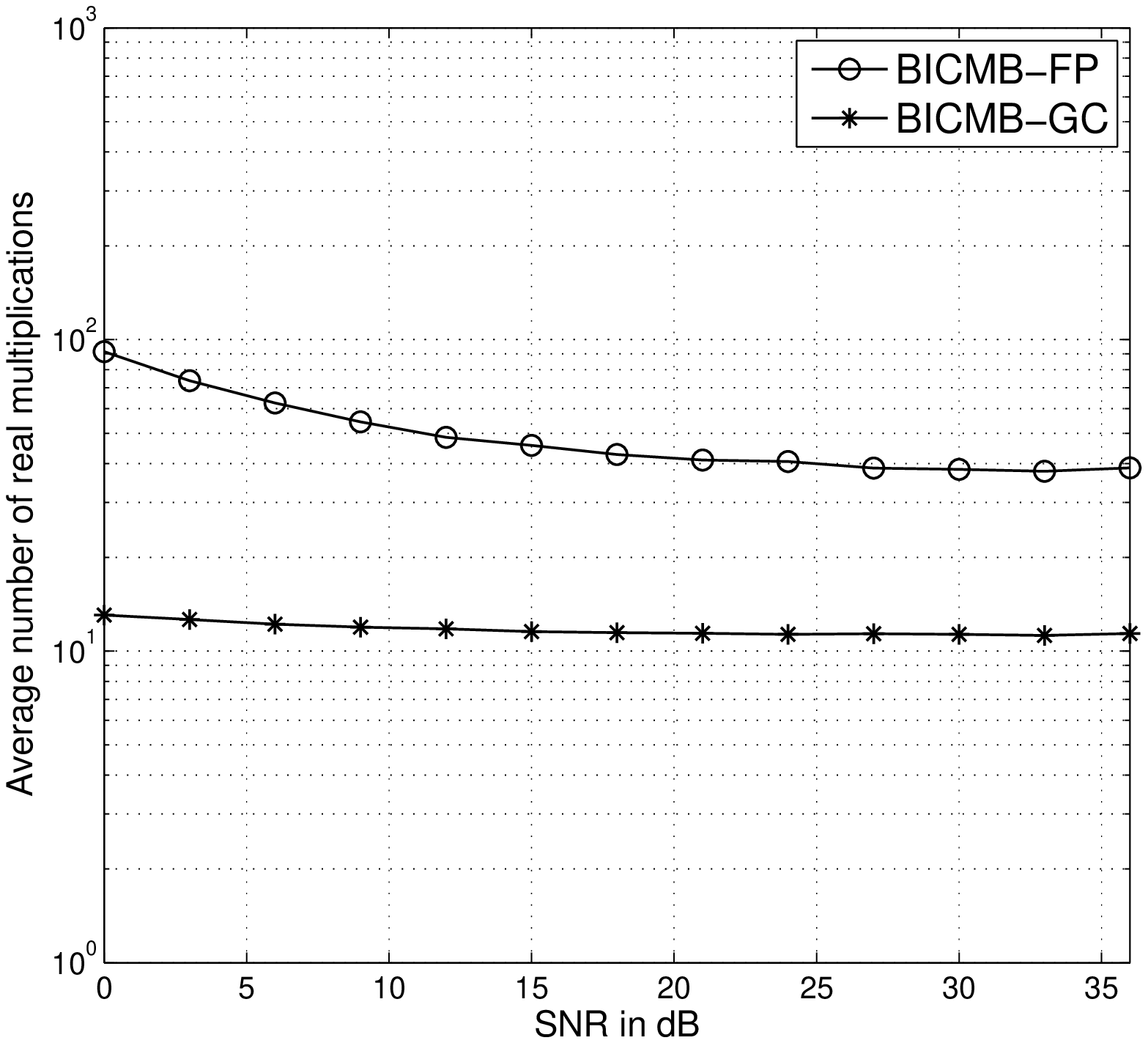}
\caption{Average number of real multiplications vs. SNR of BICMB-FP and BICMB-GC for $R_c=2/3$, $2\times2$ systems using $64$-QAM.}
\label{fig:complexity_bicmb-gc}
\end{figure}

\begin{figure}[!t]
\centering \includegraphics[width = \sizefig\linewidth]{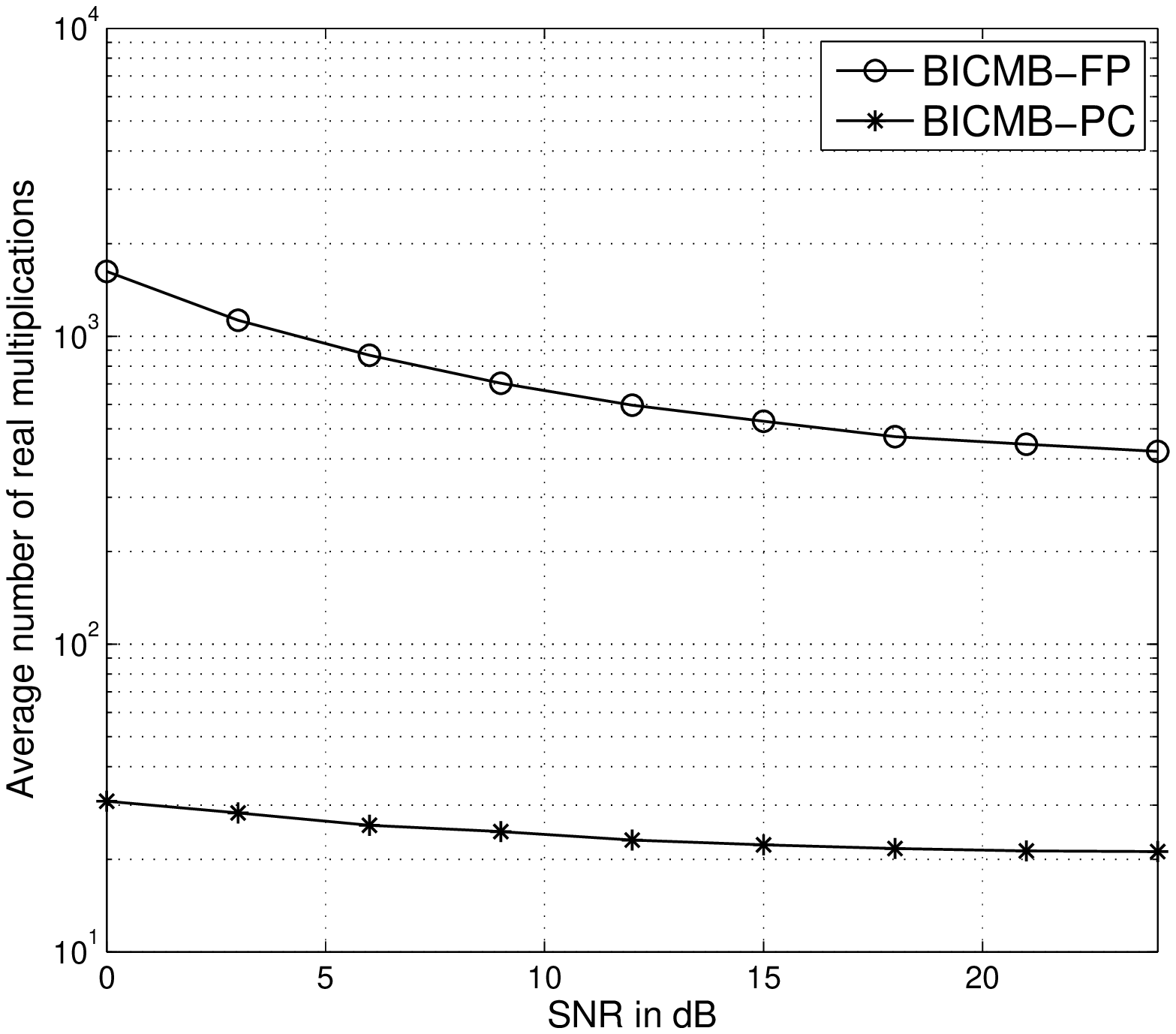}
\caption{Average number of real multiplications vs. SNR of BICMB-FP and BICMB-PC for $R_c=4/5$, $4\times4$ systems using $16$-QAM.}
\label{fig:complexity_bicmb-pc}
\end{figure}

\fi

\subsection{Discussion} \label{subsec:Results_Discussions}
As presented above, both PCMB and BICMB-PC have advantage in decoding complexity. However, both of them require the knowledge of CSIT, which is usually partial and imperfect in practice due to the bandwidth limitation and the channel estimation errors, respectively. Recently, limited CSIT feedback techniques have been introduced to achieve a performance close to the perfect CSIT case for both uncoded and coded SVD-based beamforming systems \cite{Narula_EUSI}, \cite{Xia_TB_LRF}, \cite{Love_LF_SM}, \cite{Sengul_BICMB_ICSIT}. For these techniques, a codebook of precoding matrices is known both at the transmitter and receiver. The receiver selects the precoding matrix that satisfies a desired criterion, and only the index of the precoding matrix is sent back to the transmitter. In practice, similar techniques can be applied to PCMB and BICMB-PC. On the other hand, the performance of SVD-based MIMO systems was investigated with the channel estimation errors in \cite{Au_SVD_CE}. It was shown that the performance of SVD-based MIMO systems is sensitive to the channel estimation errors. Nevertheless, space-time coding is in fact a way to improve the performance of the beamforming technique with imperfect feedback. The reason is that the spatial diversity of space-time coding, which is independent of CSIT, becomes dominant when the quality of CSI is low, and most performance gains come from the spatial diversity \cite{Jafarkhani_STC}. Note that both PCMB and BICMB-PC belong to that category. 

In this paper, the antenna configuration of $N_t=N_r=S=D$ is considered for both PCMB and BICMB-PC to make the number of transmit and receive antennas equal to PSTBC. In fact, other antenna configurations are also valid as long as $D=S\leq\min\{N_t,N_r\}$. More antennas result in greater singular values of subchannels used to transmit PSTBC, which leads to performance increase. Similarly, when the channel is frequency-selective instead of flat fading, Orthogonal Frequency Division Multiplexing (OFDM) can be applied to increase the diversity of coded SVD-based beamforming technique \cite{Akay_BICMB}, which results in performance enhancement as well. 

%% file: Conclusion.tex
\section{Conclusion} \label{sec:Conclusion}

In this paper, two novel techniques, PCMB and BICMB-PC, are presented. PCMB and BICMB-PC combine PSTBCs with uncoded and coded multiple beamforming, respectively. As a result, PCMB achieves full diversity, full multiplexing, and full rate at the same time. The main advantage of PCMB compared to PC and FPMB is that it provides significantly lower decoding complexity than PC and FPMB, respectively, in dimensions $2$ and $4$. Although the complexity gains result in performance loss, it is negligible for dimension $2$ and modest in dimension $4$. Similarly, BICMB-PC achieves both full diversity and full multiplexing, and its BER performance is almost the same as BICMB-FP. The advantage of BICMB-PC is that it offers much lower decoding complexity than BICMB-FP in dimensions $2$ and $4$. Therefore, BICMB-PC can be applied to replace the precoded part of BICMB-PP to reduce the decoding complexity. The performance investigation for limited feedback and frequency selective channels are considered as future work. 

%% file: Acknowledgement.tex
\section*{Acknowledgement}
The authors would like to thank the editor and the anonymous reviewers whose valuable comments improved the quality of the paper. 

%% file: Appendix.tex
\appendix[Proof of real-valued $\mathbf{R}$ matrix for PCMB in dimension $4$] \label{appendix}
The generation matrix  $\mathbf{G}$ for PSTBC in dimension $4$ can be found in \cite{Oggier_PSTBC}. Let $\mathbf{f}_v$ denote the $v$th column of $\mathbf{\Lambda G}$, and let $f_{u,v}$ denote the $u$th element of $\mathbf{f}_v$ with $u,v \in \{1, \ldots, 4\}$, then
\begin{equation}
\begin{split}
&f_{u,1} = \frac{1}{\sqrt{15}}\lambda_u [1+i(-3+\theta_u^2)], \\
&f_{u,2} = \frac{1}{\sqrt{15}}\lambda_u [\theta_u+i(-3\theta_u+\theta_u^3)],\\
&f_{u,3} = \frac{1}{\sqrt{15}}\lambda_u [(-3\theta_u+\theta_u^3)+i(-1+4\theta_u-\theta_u^3)],\\
&f_{u,4} = \frac{1}{\sqrt{15}}\lambda_u [(-1-3\theta_u+\theta_u^2+\theta_u^3)+i],
\label{eq:Lambda_G_4}
\end{split}
\end{equation}
where
\begin{align*}
&\theta_1 = 2\cos(4\pi/15), \qquad
&\theta_2 = 2\cos(2\pi/15), \\
&\theta_3 = 2\cos(16\pi/15),\qquad
&\theta_4 = 2\cos(8\pi/15).
\end{align*}
Note that $\theta_u^4-\theta_u^3-4\theta_u^2+4\theta_u+1 = 0$ for $u \in \{1, \ldots, 4\}$ \cite{Oggier_PSTBC}.

The nonzero elements of the diagonal matrix $\mathbf{R}$ are calculated as
\begin{equation}
\begin{split}
&r_{(1,1)} = \| \mathbf{f}_1 \|, \\
&r_{(1,2)} = \frac{\mathbf{f}_2^H\mathbf{f}_1}{\| \mathbf{f}_1 \|} =  \frac{\sum_{u=1}^4(-1+5\theta_u-\theta_u^3)}{15 \| \mathbf{f}_1 \|}, \\
&r_{(1,3)} = \frac{\mathbf{f}_3^H\mathbf{f}_1}{\| \mathbf{f}_1 \|} = \frac{\sum_{u=1}^4(4-10\theta_u-\theta_u^2+3\theta_u^3)}{15 \| \mathbf{f}_1 \|}, \\
&r_{(1,4)} = \frac{\mathbf{f}_4^H\mathbf{f}_1}{\| \mathbf{f}_1 \|} = \frac{\sum_{u=1}^4(-4-3\theta_u+2\theta_u^2+\theta_u^3)}{15 \| \mathbf{f}_1 \|}, \\
&r_{(2,2)} = \left\| \mathbf{f}_2-\frac{ \mathbf{f}_1^H\mathbf{f}_2} { \| \mathbf{f}_1 \| ^2 } \mathbf{f}_1 \right\|,\\
&r_{(2,3)} = \frac{\mathbf{f}_3^H\mathbf{f}_2}{\| \mathbf{f}_2 \|} = \frac{\sum_{u=1}^4(-3-8\theta_u+2\theta_u^2+2\theta_u^3)}{15 \| \mathbf{f}_2 \|}, \\
&r_{(2,4)} = \frac{\mathbf{f}_4^H\mathbf{f}_2}{\| \mathbf{f}_2 \|} = \frac{\sum_{u=1}^4(-1-8\theta_u+\theta_u^2+3\theta_u^3)}{15 \| \mathbf{f}_2 \|}, \\
&r_{(3,3)} = \left\| \mathbf{f}_3-\frac{ \mathbf{f}_1^H\mathbf{f}_3} { \| \mathbf{f}_1 \| ^2 } \mathbf{f}_1 - \frac{ \mathbf{f}_2^H\mathbf{f}_3} { \| \mathbf{f}_2 \| ^2 } \mathbf{f}_2 \right\|, \\
&r_{(3,4)} = \frac{\mathbf{f}_4^H\mathbf{f}_3}{\| \mathbf{f}_3 \|} = \frac{\sum_{u=1}^4(-1+5\theta_u-\theta_u^3)}{15 \| \mathbf{f}_3 \|}, \\
&r_{(4,4)} = \left\| \mathbf{f}_4-\frac{ \mathbf{f}_1^H\mathbf{f}_4} { \| \mathbf{f}_1 \| ^2 } \mathbf{f}_1 - \frac{ \mathbf{f}_2^H\mathbf{f}_4} { \| \mathbf{f}_2 \| ^2 } \mathbf{f}_2 - \frac{ \mathbf{f}_3^H\mathbf{f}_4} { \| \mathbf{f}_3 \| ^2 } \mathbf{f}_3\right\|.
\label{eq:R_elements_4}
\end{split}
\end{equation}
Based on (\ref{eq:R_elements_4}), the $\mathbf{R}$ matrix is real-valued for PCMB in dimension $4$, which is due to the special property of the $\mathbf{G}$ matrix.